\documentclass[usegraphicx,useAMS,usenatbib]{mn2e}

\usepackage{aas_macros}
\usepackage{amsmath,amssymb,amsfonts,amscd}

\usepackage[svgnames]{xcolor}
\usepackage{xspace}

\newcommand{\ie}{i.e.\xspace}

\newcommand{\eg}{e.g.\xspace}
\newcommand{\viz}{viz.\xspace}
\newcommand{\wrt}{w.r.t.\xspace}

\newcommand{\msol}{M_{\odot}}

\newcommand{\Msol}{M_{\odot}}

\newcommand{\zehn}[1]{10^{#1}}
\newcommand{\zehnh}[2]{{#1} \times 10^{#2}}

\newcommand{\ms}{\textrm{ms}}

\newcommand{\km}{\textrm{km}}
\newcommand{\cm}{\textrm{cm}}

\newcommand{\cmcmis}{\textrm{cm}^{2} \, \textrm{s}^{-1}}
\newcommand{\erg}{\textrm{erg}}
\newcommand{\ergs}{\textrm{erg} \, \textrm{s}^{-1}}
\newcommand{\sek}{\textrm{s}}
\newcommand{\isek}{\textrm{s}^{-1}}

\newcommand{\gccm}{\textrm{g\,cm}^{-3}}

\newcommand{\Gauss}{\textrm{G}}

\newcommand{\MeV}{\textrm{MeV}}

\newcommand{\modl}[1]{model \texttt{#1}}
\newcommand{\modls}[1]{models \texttt{#1}}
\newcommand{\Modl}[1]{Model \texttt{#1}}
\newcommand{\Modls}[1]{Models \texttt{#1}}
\newcommand{\modelname}[1]{\texttt{#1}}

\newcommand{\nusps}{neutrinospheres\xspace}

\newcommand{\panel}[1]{\textit{#1}}
\newcommand{\ppns}{\textsc{pns}}
\newcommand{\Erot}{\mathcal{T}}
\newcommand{\Frot}{\mathcal{F}}
\newcommand{\Erotpns}{\Erot_{\ppns}}
\newcommand{\Frotpns}{\Frot_{\ppns}}
\newcommand{\Mej}{\mathcal{M}_{\rm ej}}

\newcommand{\tauAlf}{\tau_{\textrm{Alf}}}
\newcommand{\tauadv}{\tau_{\textrm{adv}}}
\newcommand{\tauhtg}{\tau_{\textrm{heat}}}
\newcommand{\tautau}{\tauadv / \tauhtg}
\newcommand{\tautaum}{\tauadv / \tauAlf}

\newcommand{\tpb}{t_{\textrm{pb}}}

\newcommand{\eej}{E_{\textrm{ej}}}
\newcommand{\mej}{M_{\textrm{ej}}}

\newcommand{\MPNS}{M_{\ppns}}

\newcommand{\igain}{{\textrm{gain}}}

\newcommand{\MPNSm}[1]{M_{\ppns}^{\modelname{#1}}}

\newcommand{\figref}[1]{Fig.\,\ref{#1}}
\newcommand{\tabref}[1]{Tab.\,\ref{#1}}
\newcommand{\secref}[1]{Sect.\,\ref{#1}}

\newcommand{\nth}[1]{${#1}^{\mathrm{th}}$}

\newcommand{\Alfven}{Alfv{\'e}n\xspace}

\defcitealias{Obergaulinger__2021__mnras__MagnetorotationalCoreCollapseofPossibleGRBProgenitorsIII.ThreeDimensionalModels}{Paper III}
\defcitealias{Aloy__2021__MonthlyNoticesoftheRoyalAstronomicalSociety__MagnetorotationalCoreCollapseofPossibleGRBProgenitorsII.FormationofProtomagnetarsandCollapsars}{Paper II}
\defcitealias{Obergaulinger_Aloy__2020__mnras__MagnetorotationalCoreCollapseofPossibleGRBProgenitorsIExplosionMechanisms}{Paper I}
\defcitealias{Aguilera-Dena_et_al__2018__apj__RelatedProgenitorModelsforLong-durationGamma-RayBurstsandTypeIcSuperluminousSupernovae}{AD18}
\defcitealias{Kasen_Bildsten__2010__apj__SupernovaLightCurvesPoweredbyYoungMagnetars}{KB10}

\newcommand{\AguiDet}{\citetalias{Aguilera-Dena_et_al__2018__apj__RelatedProgenitorModelsforLong-durationGamma-RayBurstsandTypeIcSuperluminousSupernovae}\xspace}

\begin{document}

\title[MHD core collapse across the mass range] {Magnetorotational core collapse of
  possible GRB progenitors.  IV. A wider range of progenitors}

\author[Obergaulinger \& Aloy]{
  M.~Obergaulinger$^{1,2}$, M.\'A.~Aloy$^{1,3}$
  \\
  $^1$ Departament d{\'{}}Astronomia i Astrof{\'i}sica, Universitat de
  Val{\`e}ncia, 
  C/ Dr.~Moliner, 50, 46100 Burjassot, Spain
  \\
  $^{2}$ Institut f{\"u}r Kernphysik, Technische Universit{\"a}t
  Darmstadt, Schlossgartenstra{\ss}e 2, 64289 Darmstadt, Germany
  \\
  $^{3}$ Observatori Astronòmic, Universitat de València, 46980 Paterna, Spain
}

\label{firstpage}
\pagerange{\pageref{firstpage}--\pageref{lastpage}}
\maketitle

\begin{abstract}
  The final collapse of the cores of massive stars can lead to a wide
  variety of outcomes in terms of electromagnetic and kinetic
  energies, nucleosynthesis, and remnants.  The connection of this
  wide spectrum of explosion and remnant types to the properties of
  the progenitors remains an open issue.  Rotation and magnetic fields
  in Wolf-Rayet stars of subsolar metallicity may explain extreme
  events such as superluminous supernovae and gamma-ray bursts powered
  by proto-magnetars or collapsars. Continuing numerical studies of
  magnetorotational core collapse including detailed neutrino physics,
  we focus on progenitors with zero-age main-sequence masses in the
  range between 5 and 39 solar masses. The pre-collapse stars are one
  dimensional models employing prescriptions for the effects of
  rotation and magnetic fields. Eight of the ten stars we consider
  being the results of chemically homogeneous evolution due to
  enhanced rotational mixing
  \citep[][]{Aguilera-Dena_et_al__2018__apj__RelatedProgenitorModelsforLong-durationGamma-RayBurstsandTypeIcSuperluminousSupernovae}. All
  but one of them produce explosions driven by neutrino heating (more
  likely for low mass progenitors up to 8 solar masses) and
  non-spherical flows or by magnetorotational stresses (more frequent
  above 26 solar masses). In most of them and for the one
  non-exploding model, ongoing accretion leads to black-hole
  formation. Rapid rotation makes a subsequent collapsar activity
  plausible. Models not forming black holes show proto-magnetar driven
  jets. Conditions for the formation of nickel are more favourable in
  magneto-rotationally driven models, though our rough estimates fall
  short of the requirements for extremely bright events if these are
  powered by radioactive decay. However, approximate light curves of
  our models suggest that a proto-magnetar or black hole spin-down may
  fuel luminous transients (with peak luminosities $\sim
  \zehn{43\ldots 44}\,\erg$).
\end{abstract}


\section{Introduction}
\label{Sek:Intro}

The core collapse of a star with more than $\sim 8 \, \Msol$ after the
end of its hydrostatic burning phases initiates a sequence of
processes leading to one of several quite different outcomes. The star
may explode as a core-collapse supernova (CCSN) or not explode at all,
while the long-term central compact remnant can either be a black hole
(BH) or a neutron star (NS). Furthermore, in special cases, the
dynamics of the compact remnant may produce suitable conditions for an
even more luminous high-energy transient, namely a long gamma-ray
burst (GRB).  These explosions are ultimately powered by the
gravitational binding energy liberated during the collapse and the
subsequent accretion of matter onto the core.  While this energy
source is universal, there are different mechanisms for converting a
fraction of it to the kinetic and internal energy of the ejecta of the
explosion.  Hence, the evolution depends on the pre-collapse state of
the progenitor star and on a multitude of complex, interconnected
processes occurring during and after the collapse and the birth of a
proto-neutron star (PNS).

The detection of neutrinos from supernova SN 1987A, a large number of
increasingly detailed multi-dimensional simulations, and the
comparison of their results to electromagnetic observations of CCSNe
and young supernova remnants (SNRs) offer confirmation for the
standard neutrino-driven mechanism \citep[for a review,
see][]{Janka__2012__ARNPS__ExplosionMechanismsofCore-CollapseSupernovae}.
This scenario, based on neutrinos depositing energy and on non-radial
fluid instabilities in the hot bubble surrounding the PNS, is able to
explain the bulk of the observed CCSNe without requiring special
conditions for the progenitor.

However, it remains difficult to account for the most extreme events
such as hypernovae \citep[HNe;][]{Iwamoto_1998Natur.395..672}, \ie, the
most energetic class of CCSNe, or long GRBs characterized by
collimated, relativistic outflows.  Since these events constitute only
a minor fraction of the entire population, they have been connected to
special progenitor classes, in particular stars with high rotation
rates.  In these cases, magnetic fields may play a crucial role in
launching the explosion by tapping into the rotational energy.
Additionally, rotation and magnetic fields have been invoked to
explain the extraordinary brightness of super-luminous supernovae
\citep[SLSNe;
e.g.][]{GalYam_2012Sci...337..927,Moriya_et_al__2018__ssr__SuperluminousSupernovae, 
Gal-Yam_2019ARA&A..57..305, Inserra_2019NatAs...3..697}.

The basic building blocks of magnetorotational models are either a
fast spinning, strongly magnetized PNS, a so-called proto-magnetar
(PM; \cite{Kasen_Bildsten__2010__apj__SupernovaLightCurvesPoweredbyYoungMagnetars} -\citetalias{Kasen_Bildsten__2010__apj__SupernovaLightCurvesPoweredbyYoungMagnetars} hereafter-;
\cite{Woosley__2010__apjl__Bright_SNe_from_Magnetar_Birth,Metzger_et_al__2011__mnras__Theprotomagnetarmodelforgamma-raybursts, Nicoll_2017ApJ...850...55}), whose spin-down drives the explosion
(HNe or GRBs) or heats the ejecta (SLSNe), or accretion of rapidly
rotating matter onto a BH magnetically powering relativistic jets
\citep[collapsar model for
GRBs;][]{MacFadyen__2001__apj__Supernovae_Jets_and_Collapsars,Obergaulinger_Aloy__2017__mnras__Protomagnetarandblackholeformationinhigh-massstars}.

Whether the strong magnetic fields required for these types of
explosions can be reached depends on various processes amplifying them
in the collapsed core and on the pre-collapse configuration.
Therefore, models for this class of explosions should be based on
progenitor stars that include the effects of rotation and magnetic
fields in a self-consistent manner.  However, the computational costs
for covering the entire pre-collapse life of a star up to collapse
restrict stellar evolution calculations to the assumption of spherical
symmetry, thus allowing for an inclusion of these effects only in
approximate ways.  Among the most advanced such approximations is the
model for a dynamo in convectively stable layers of stars proposed by
\cite{Spruit__2002__AA__Dynamo} which has been incorporated into
spherically symmetric stellar evolution models to produce pre-collapse
models
\citep[\eg,][]{Heger_et_al__2005__apj__Presupernova_Evolution_of_Differentially_Rotating_Massive_Stars_Including_Magnetic_Fields}.
\cite{Woosley_Heger__2006__apj__TheProgenitorStarsofGamma-RayBursts}
presented a set of potential progenitors for GRBs that have found wide
application in numerical investigations of stellar core collapse and
its aftermath.  Among those studies, we refer to the previous articles
in this series
\citep[][\citetalias{Obergaulinger_Aloy__2020__mnras__MagnetorotationalCoreCollapseofPossibleGRBProgenitorsIExplosionMechanisms,Aloy__2021__MonthlyNoticesoftheRoyalAstronomicalSociety__MagnetorotationalCoreCollapseofPossibleGRBProgenitorsII.FormationofProtomagnetarsandCollapsars,Obergaulinger__2021__mnras__MagnetorotationalCoreCollapseofPossibleGRBProgenitorsIII.ThreeDimensionalModels},
hereafter]{Obergaulinger_Aloy__2020__mnras__MagnetorotationalCoreCollapseofPossibleGRBProgenitorsIExplosionMechanisms,Aloy__2021__MonthlyNoticesoftheRoyalAstronomicalSociety__MagnetorotationalCoreCollapseofPossibleGRBProgenitorsII.FormationofProtomagnetarsandCollapsars,Obergaulinger__2021__mnras__MagnetorotationalCoreCollapseofPossibleGRBProgenitorsIII.ThreeDimensionalModels}
evolving various variations of two progenitors in axisymmetry as well
as three dimensions including all the physics relevant during the
formation of the PNS and the subsequent evolution until explosion (or
lack thereof). We note that many other works took a different approach
and simulated the propagation of relativistic jets in progenitors of
this group under the assumption that the cores develop the conditions
necessary for producing a GRB \citep[for an overview, see,
\eg,][]{Corsi__2021__NewAstronomyReviews__GammaRayBurstJetsinSupernovae}.

More recently,
\cite{Aguilera-Dena_et_al__2018__apj__RelatedProgenitorModelsforLong-durationGamma-RayBurstsandTypeIcSuperluminousSupernovae}
\citepalias{Aguilera-Dena_et_al__2018__apj__RelatedProgenitorModelsforLong-durationGamma-RayBurstsandTypeIcSuperluminousSupernovae}
as well as
\cite{AguileraDena__2020__TheAstrophysicalJournal__PrecollapsePropertiesofSuperluminousSupernovaeandLongGammaRayBurstProgenitorModels}
have produced a sequence of rotating, magnetized pre-collapse models
for Wolf-Rayet stars marked by strong stellar winds.  The resulting
loss of the outer layers is, besides rotation and magnetic fields,
another important ingredient in models for the aforementioned extreme
explosions.  It is, however, accompanied by a loss of angular
momentum, which might reduce the prospects of retaining enough
rotational energy for such an explosion.  Nevertheless, these stars
undergo a chemically homogeneous evolution (CHE) due to an adjustment
of the mixing of gas between different layers induced by rotation and
end their lifes with high angular velocities.  \AguiDet used an
approximation to predict the possible outcome of the core collapse of
their models, finding several candidates for SLSNe and for GRBs driven
by PMs and collapsars.

As the studies on magnetorotational core collapse summarized above
indicate, the models of \AguiDet in principle possess the right
properties to produce SLSNe and GRBs.  Thus, a closer investigation by
self-consistent simulations of the collapse and the subsequent phases
is needed, and this is the motivation of this paper.  We complement
the set of CHE models with \modl{16TI} from
\cite{Woosley_Heger__2006__apj__TheProgenitorStarsofGamma-RayBursts}.
This star has an initial mass of $M_{\textsc{ZAMS}} = 16 \, \Msol$ and
sub-solar metallicity ($Z=0.1 \, Z_\odot$).  \modelname{16TI} is a
plausible collapsar-forming progenitor due to its structure and in
particular its fast rotation. It has been used in several studies of
the propagation of jets launched by an BH-accretion-disc system
through the stellar envelope.  However, previous work
\citep[\eg,][]{Lazzati__2009__TheAstrophysicalJournal__VeryHighEfficiencyPhotosphericEmissioninLongDurationGammaRayBursts,Lazzati_et_al__2012__apj__UnifyingtheZooofJet-drivenStellarExplosions,
  Hayakawa_Maeda__2018__apj__ACollapsarModelwithDiskWindImplicationsforSupernovaeAssociatedwithGammaRayBursts,
  Aloy__2018__MonthlyNoticesoftheRoyalAstronomicalSociety__OntheExistenceofaLuminosityThresholdofGRBJetsinMassiveStars},
assumes that the core has formed such a BH-accretion-disc structure.
The central regions, including the BH-torus system or any other type
of compact remnant, are excised and a jet injected through a suitably
chosen inner boundary is evolved in the framework of relativistic
(magneto-)hydrodynamics.  By simulating the phase prior to that stage,
we complement these studies and explore the possibility of this star
actually satisfying the conditions for launching a collapsar jet in
the first place.

We combine magnetohydrodynamics (MHD) and a two-moment neutrino
transport to follow the evolution of eight CHE progenitor models
between $5$ and $39 \,\Msol$, along with \modl{16TI}.  Our goal is to
check the predictions regarding the evolution of the stars.  To this
end, long simulation times and a large number of models are required.
For this reason, we restrict ourselves at this point to
two-dimensional models and defer detailed three-dimensional models to
a later stage.

This article is organized as follows: \secref{Sek:Num} briefly introduces the
physics included in the simulations and the numerical methods,
\secref{Sek:Init} summarizes the progenitor models selected in this
study, \secref{Sek:Res} presents the simulation results,
\secref{Sek:espekulatius} roughly estimates the observable signal of the
models, and \secref{Sek:Concl} ends the article with concluding remarks.

\section{Input physics and numerics}
\label{Sek:Num}

Methodologically, the present study follows our previous work on
stellar core collapse
(\cite{Obergaulinger_et_al__2014__mnras__Magneticfieldamplificationandmagneticallysupportedexplosionsofcollapsingnon-rotatingstellarcores,Obergaulinger_Aloy__2017__mnras__Protomagnetarandblackholeformationinhigh-massstars}; \citetalias{Obergaulinger_Aloy__2020__mnras__MagnetorotationalCoreCollapseofPossibleGRBProgenitorsIExplosionMechanisms}).
We employed the simulation code Alcar
\citep{Just_et_al__2015__mnras__Anewmultidimensionalenergy-dependenttwo-momenttransportcodeforneutrino-hydrodynamics}
for solving the combined systems of equations of special relativistic
MHD equations and spectral two-moment neutrino transport.  The gas is
described by the SFHo equation of state (EOS) of
\cite{Steiner_et_al__2013__apj__Core-collapseSupernovaEquationsofStateBasedonNeutronStarObservations}
applied at densities $\rho > \rho_{\mathrm{EOS}} = 6000 \, \gccm$ and
an EOS accounting for a mixture of photons, electrons and positrons,
and nucleons below this threshold.  The nuclear composition is
determined from nuclear statistical equilibrium (NSE) and using the
flashing scheme of \cite{Rampp_Janka__2002__AA__Vertex} in the
high-density and low-density regimes, respectively.  The self-gravity
of the gas is modelled by version 'A' of the  TOV potential of
\cite{Marek_etal__2006__AA__TOV-potential}, which yields a very
close approximation to general relativity in terms of, \eg, the
evolution of the density profiles of PNSs and the maximum mass of stable
cold neutron stars.

We evolve neutrinos in the two-moment transport scheme consisting of a
set of balance equations for their energy (\nth{0} moment) and
momentum ($1^{\mathrm{st}}$ moment) densities closed by a local
algebraic expression for the pressure tensor ($2^{\mathrm{nd}}$
moment).  One pair of these equations is solved for each neutrino
species (electron neutrinos, electron anti-neutrinos, and heavy-lepton
neutrinos) and for each one of several bins in the space of neutrino
energies.  The energy bins are coupled by velocity-dependent and
gravitational terms in the $\mathcal{O} (v/c)$-plus method of
\cite{Endeve_et_al__2012__ArXive-prints__ConservativeMomentEquationsforNeutrinoRadiationTransportwithLimitedRelativity}.
Neutrinos and matter interact via the most important reactions:
nucleonic and nuclear absorption, emission, and scattering, inelastic
scattering off electrons, and the pair processes of electron-positron
annihilation and nucleonic bremsstrahlung.

The simulations are run in spherical coordinates assuming axisymmetry.
We set up grids with $n_r = 480$ radial zones distributed
logarithmically up to an outer radius of $R_{\mathrm{out}} =
\zehnh{7}{10} \, \cm$ and $n_{\theta} = 128$ zones in angle.  In
energy, a logarithmic grid of $n_{\epsilon} = 10$ bins with a  maximum
energy of $\epsilon_{\mathrm{max}} = 240 \, \MeV$ is used.

\section{Initial models}
\label{Sek:Init}

Our set of models includes a progenitor employed in our previous
studies, \viz a star of an initial mass of
$M_{\textsc{ZAMS}} = 35 \, \Msol$ and sub-solar metallicity
($Z=0.1 Z_\odot$) evolved to the pre-collapse stage with rotation and
magnetic fields by
\cite{Woosley_Heger__2006__apj__TheProgenitorStarsofGamma-RayBursts}.
Among the different versions of the models based on the progenitor
\modelname{35OC}, we only use the one with profiles of rotation and
magnetic field taken from the stellar evolution calculations as a
comparison model here.  The model was named \modelname{35OC-RO} in the
previous articles of this series; for the sake of brevity, we refer to
it as \modl{35OC} in the following.

All the other initial conditions of our simulations are given by eight
stellar models computed by \AguiDet.  These are models for rapidly
rotating Wolf-Rayet stars with magnetic fields at subsolar metallicity
of $Z = 0.02 \, Z_{\odot}$ evolved in spherical symmetry using the
Modules for Experiments in Stellar Astrophysics (MESA) code
\citep{Paxton__2011__ApJS__ModulesforExperimentsinStellarAstrophysicsMESA,Paxton__2013__ApJS__ModulesforExperimentsinStellarAstrophysicsMESAPlanetsOscillationsRotationandMassiveStars,Paxton__2015__ApJS__ModulesforExperimentsinStellarAstrophysicsMESABinariesPulsationsandExplosions,Paxton__2018__ApJS__ModulesforExperimentsinStellarAstrophysicsMESAConvectiveBoundariesElementDiffusionandMassiveStarExplosions}.
The magnetic fields are evolved following the Taylor-Spruit dynamo
\citep{Heger_et_al__2005__apj__Presupernova_Evolution_of_Differentially_Rotating_Massive_Stars_Including_Magnetic_Fields},
i.e., the treatment of magnetic fields through the stellar evolution
is effectively the same as in models \modelname{35OC} and
\modelname{16TI}.  The stars are set up with an angular velocity rate
close to the critical rotation rate.  The fast rotation leads to
efficient mixing across different layers of the star. Thereby, these
models undergo a chemically homogeneous evolution (CHE).  \AguiDet
computed two sets of models, one with the standard sets of parameters
and another one with an increased rate of rotational mixing (series
B).  It is from the latter that we select our pre-collapse models.

At the end of their hydrostatic evolution, the stars of series B have
lost their H envelope and most of the He layer.  They retain a large
amount of angular momentum and possess magnetic fields whose strength
depends on the specific progenitor.  These properties make them
possible progenitors for type Ic SLSNe or GRBs.  \AguiDet explored the
types of explosions to be expected as follows:
\begin{itemize}
\item The stars with lowest initial masses ($M_{\textsc{ZAMS}}= 5$, $8$, and
  $13 \, \msol$) may explode as type Ic SLSNe in which the excess
  luminosities are caused mostly by energy input from a central PM. 
\item The stars with higher masses fulfil the expectations for forming
  a GRB, in particular in terms of the angular momentum in the core.
  \AguiDet suggest that the ones in an intermediate mass range (models
  with $17$ and $20 \, \msol$) might avoid BH formation and thus yield
  to PM-powered GRBs.
\item More massive stars (in our sample, the ones with $26$, $30$, and
  $39 \, \msol$) might end their lifes as collapsars.
\item Though we do not include them in our sample, we note for the
  sake of completeness that \AguiDet also consider more massive stars
  that might, after undergoing pulsational pair instability, explode
  as GRBs or HNe and, by interaction with the stellar winds evolve
  into SLSNe.
\end{itemize}

In the process of mapping the stellar models at the onset of collapse
onto our 2d grid, we have to convert the radial profiles of the
poloidal and toroidal field components into the three vector
components of the magnetic field, $(b^r, b^{\theta}, b^{\phi})$, as a
function of radius and latitude.  Proceeding in the same way as in
\cite{Obergaulinger_Aloy__2017__mnras__Protomagnetarandblackholeformationinhigh-massstars,Obergaulinger_Aloy__2020__mnras__MagnetorotationalCoreCollapseofPossibleGRBProgenitorsIExplosionMechanisms},
we assume a large-scale geometry of the field restricted to the layers
of the star which are magnetized.  The radial and lateral components,
on the one hand, and the azimuthal components, on the other hand, are
respectively computed from the poloidal and the toroidal components of
the stellar evolution model, all of them modulated by a factor
$\sin \theta$.  We note that, since the Taylor-Spruit dynamo is
designed for radiative layers of stars, the magnetic field vanishes in
convective layers by construction.

\begin{figure}
  \centering
  \includegraphics[width=\linewidth]{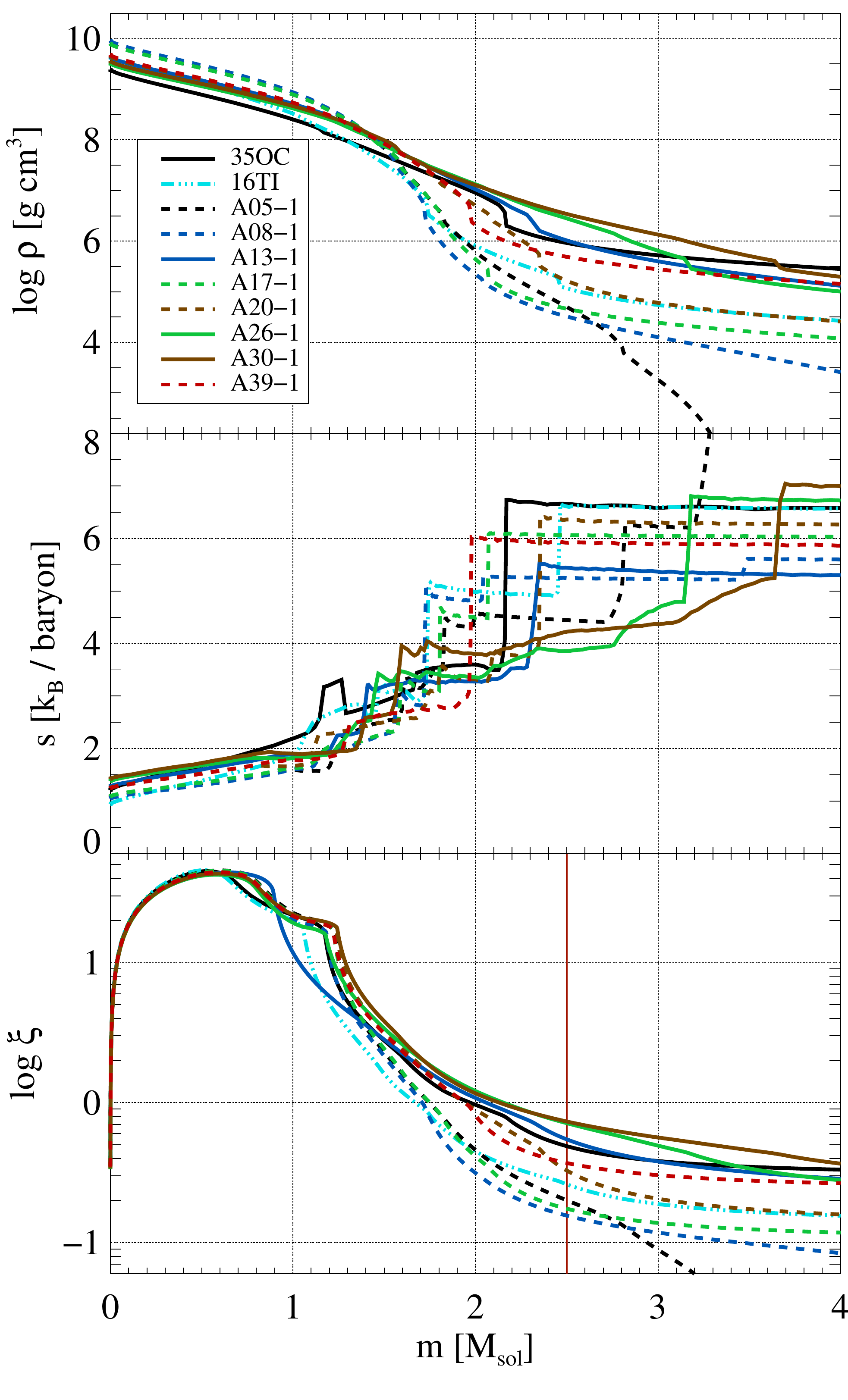}
  \caption{
    Top to bottom: profiles of density, specific entropy, and
    compactness of all progenitors at the onset of collapse. Models
    producing neutrino-driven explosions, MHD effects or fail to
    explode are displayed with dashed, solid and dash-dotted lines,
    respectively.  The vertical line in the bottom panel marks the
    mass coordinate $m=2.5M_\odot$, facilitating the comparison of the
    values of $\xi$ at that mass coordinate for each model. Here, the
    compactness $\xi$ is not evaluated at the time of bounce as it is
    the case of $\xi_{2.5}$ in Tab.\,\ref{Tab:Globerview}.  }
  \label{Fig:initmodels}
\end{figure}

\begin{figure*}
  \centering
  \includegraphics[width=\linewidth]{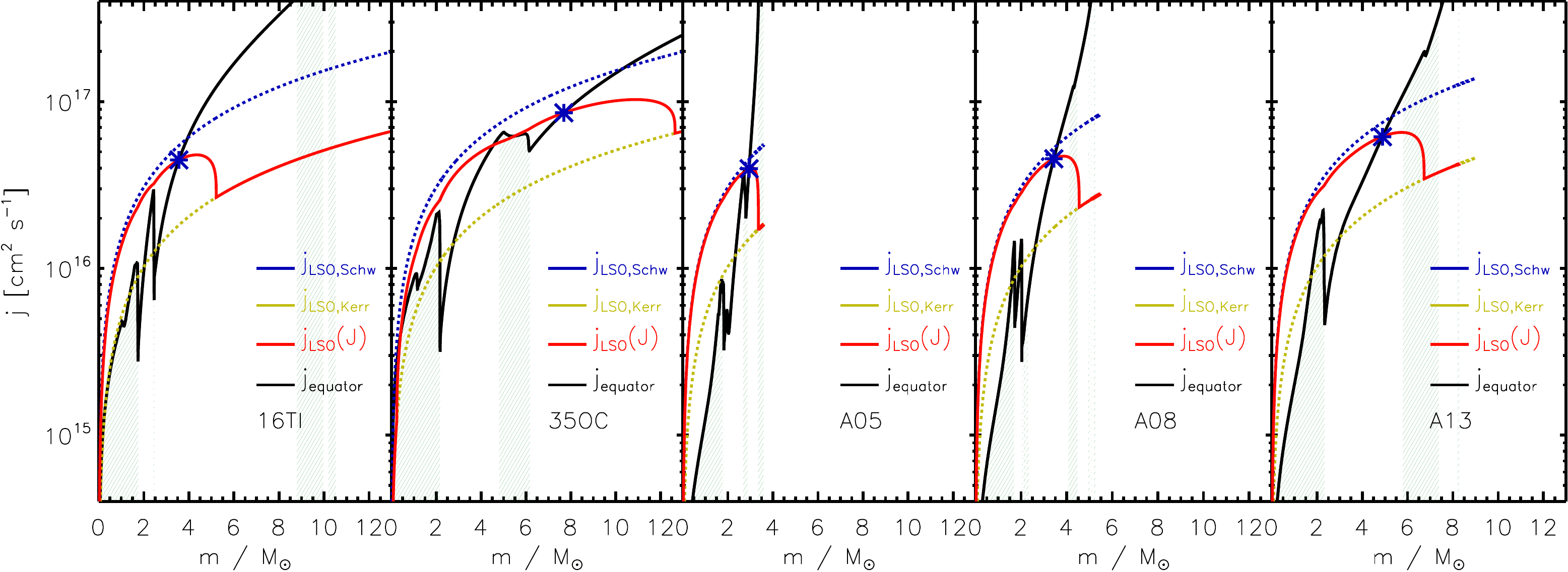}
  \includegraphics[width=1.01\linewidth]{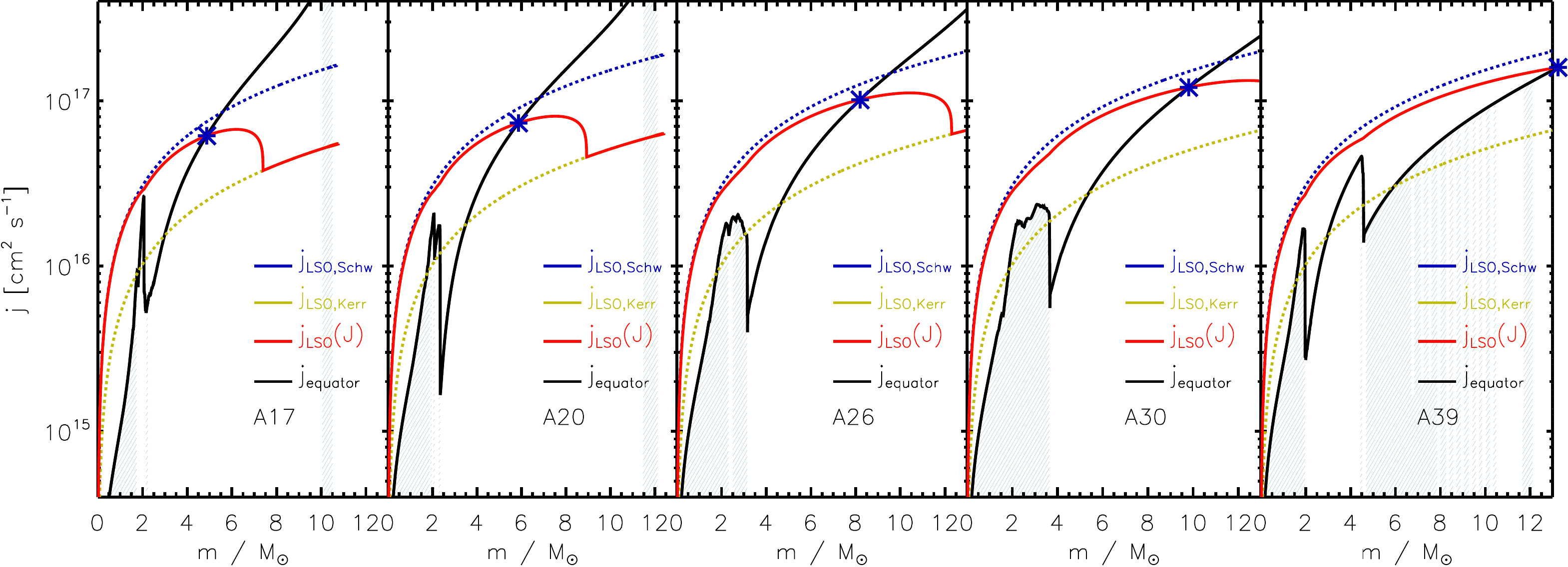}
  \caption{
    Equatorial profile of the initial specific angular momentum
    (black lines) of models as indicated in the panels. In
    each panel, the blue dashed lines denote the angular momentum
    needed to support matter at the LSO for a Schwarzschild BH,
    while the yellow dashed lines are for a Kerr BH with
    dimensionless spin $a=1$.  The red lines indicate the specific
    angular momentum at the LSO for a BH with the mass and angular
    momentum inside the displayed mass coordinate in the pre-SN
    star.  The green-hatched parts of the plot denote
    the mass shells of the pre-SN star with non-zero magnetic
    field.
  }
  \label{Fig:init-angmom}
\end{figure*}

We present the structure of the progenitors at the onset of collapse
in \figref{Fig:initmodels} and \figref{Fig:init-angmom}.  In the
former figure, the \panel{top panel} displays the density profiles of
the stars, the \panel{mid panel} with profiles of the specific
entropy gives an overview of the location of shell interfaces in the
cores, and the \panel{bottom} panel shows the compactness parameter
\citep{OConnor_Ott__2011__apj__BlackHoleFormationinFailingCore-CollapseSupernovae},
\begin{align}
\xi (m) = \left( m/\Msol\right) \left( 1000 \, \km / r(m) \right),
\label{eq:xi}
\end{align}
defined by the mass $m$ enclosed in a radius $r(m)$.  The latter
figure compares the profiles of specific angular momentum in the
equatorial planes of the progenitors to the specific angular momentum
of the last stable orbits for gas at a Lagrangian coordinate $m$
orbiting black holes that would form from the collapse of all the
matter inside of it.  For those comparison profiles, we assume
non-rotating and maximally rotating BHs as well as BHs whose angular
momentum is equal to that of all the layers up to the mass coordinate
$m$.  Shells whose actual specific angular momentum exceeds the ones
of the last stable orbits can be expected to form discs around BHs
rather than directly fall into them (the mass coordinate where this
happens is marked with a blue asterisk).  The fact that in all models
of \figref{Fig:init-angmom} matter above a finite mass coordinate can
form an accretion disc indicates that in all of them a collapsar could
form if a the compact remnant left after core collapse becomes a
BH. Should the post-collapse PNS become a NS in the long term, the
high specific angular momentum of these layers may limit the amount of
mass that can be accreted over the central object.  Furthermore,
layers of the stars with non-vanishing magnetic fields are shaded in
these panels.

\section{Results}
\label{Sek:Res}

\subsection{Overview}
\label{sSek:Overview}

\begin{table*}
  \centering
  \begingroup
  \setlength{\tabcolsep}{4pt} 
  \begin{tabular}{| l | ccc ccc cc | lll lll lll ll |}
    \hline\\*[-9pt]
    model & AD18 & $M_4$ & $\mu_4$  & $\mu_4M_4$ &$\mathcal{T}_4$  &
    $\lambda_{2.5}$ & $\overline{B}_{2.5}^{\,\rm p}$ & $\overline{B}_{2.5}^{\,\rm t}$ &
                                                                     $\xi_{2.5}$
                                                                     & $t_{\mathrm{f}}$ & $t_{\mathrm{exp}}$ & $M_{\mathrm{sh,e}}$ &
                                                                      kind
                                                                    &
                                                                    $E_{\mathrm{exp}}$
    & $\Frotpns$ & $\Erotpns$ 
    & remn. & $M_{\mathrm{r}}$ & $M_{\mathrm{Ni}}$
    \\ 
          & & & & & & &$[10^{8} \, \text{G}]$ & {$[10^{10} \, \text{G}]$}  & 
    & $[\sek]$ & $[\sek]$ & $[\msol]$ & & \multicolumn{3}{c}{$[10^{51} \, \erg]$}
                                  
    & & \multicolumn{2}{c|}{$[\Msol]$}
    \\ \hline
    35OC & $--$ & 2.15 & 0.20 & 0.43 &{8.61} & {0.58} & {$250$}& {$54.8$}& 0.49 & 2.6 & 0.19 & 1.9 &
                                                                       $\nu$/M
                                                                       &
                                                                         1.2
    & 5.5 & {148.5} 
    & \textbf{BH} & 2.9 & 0.048
    \\
    16TI & $--$ & 1.73 & 0.14 & 0.25  & 2.9 & {0.13} &{$24.5$} & {17.7} & 0.26 & 6.1 & $--$ & $--$
                                                                  & $--$
                                                                    &
                                                                      $--$ 
                                                                      & 1.1 & {128.7}
    & BH & 2.8 & $--$
    \\ \hline
    A05 & SLSN & 1.82 & 0.11 & 0.21  & 0.70 & {0.05} & {0.70} & {1.26} & 0.20 & 3.7 & 1.7 & 2.1 &
                                                                      $\nu$
                                                                      &
                                                                        1.9
                                                                        & 0.40 & {10.64}
    & PNS & 2.2 & 0.011
    \\ 
    A08 & SLSN & 1.72 & 0.09 & 0.16 & 1.4 & {0.06}& {0.89}& {0.55} & 0.16 & 4.8 & 3.4 & 2.1 &
                                                                     $\nu$
                                                                     &
                                                                       0.43
                                                                       & 0.54 & {18.11}
    & PNS & 2.2 & 0.003 
    \\ 
    A13 & SLSN & 2.30 & 0.19 & 0.44 & 2.5 & {0.69} & {8.26}& {4.15} & 0.55 & 2.9 & 0.39 & 2.2 &
                                                                      M
                                                                      &
                                                                        0.76
                                                                        & 0.09 & {45.74}
    & \textbf{BH} & 2.6 & 0.050
    \\ 
    A17 & PM & 2.07 & 0.05 & 0.11 & 0.76 & {0.05} & {1.42}& {0.36}& 0.17 & 3.9 & 1.7 & 2.1 &
                                                                    $\nu$
                                                                    &
                                                                      1.1
                                                                      & 2.2 & {17.45}
    & PNS & 2.1 & 0.009 
    \\ 
    A20 & PM & 2.34 & 0.08 & 0.18 & 2.2 & {0.09} & {0.57}& {0.91} & 0.32 & 4.0 & 1.2 & 2.4 &
                                                                   $\nu$
                                                                   &
                                                                     0.49
                                                                     & 2.7 & {53.58}
    & \textbf{BH} & 2.5 & 0.011
    \\ 
    A26 & {Col} & 2.76 & 0.21 & 0.58 & 2.9 & {0.74} &{3.64} & {2.53} & 0.71 & 1.5 & 0.56 & 2.5 &
                                                                    M
                                                                    &
                                                                      0.98
                                                                      & 1.5 & {83.49}
    & \textbf{BH} & 3.0 & 0.022
    \\ 
    A30 & {Col} & 2.28 & 0.33 & 0.75 & 2.1& {0.98} & {6.85} & {3.72} & 0.73 & 1.5 & 0.39 & 2.3 &
                                                                    M
                                                                    &
                                                                      1.4
                                                                      & 2.8 & {89.74}
    & \textbf{BH} & 3.0 & 0.036
    \\ 
    A39 & {Col} & 1.97 & 0.12 & 0.25 & 2.6& {0.33} & {8.87} & {4.31} & 0.37 & 3.8 & 0.51 & 2.1 &
                                                                    $\nu$/M
                                                                    &
                                                                      0.56
                                                                      & 5.8 & {41.20}
    & \textbf{BH} & 2.4 & 0.030
    \\    \hline
  \end{tabular}
  \endgroup
  \caption{%
    Overview of models: model name, evolution according to \AguiDet
    (superluminous SN, PM with the potential to produce a GRB, or
    collapsar -denoted with ``Col''-), properties of the stellar progenitor at the presupernova link, $M_4$
    (Eq.\,\eqref{eq:M4}), $\mu_4$ (Eq.\,\eqref{eq:mu4}), $\lambda_{2.5}$ (Eq.\,\eqref{eq:lambda25})
    $\mathcal{T}_4$ (the rotational energy of the matter inside
    $M_4$ in units of $[10^{48} \, \erg]$), and volume-averaged
    poloidal and toroidal magnetic filed strength within the inner
    $2.5M_\odot$. The compactness parameter
    $\xi_{2.5}:=\xi(2.5M_\odot)$ (Eq.\,\eqref{eq:xi}) is measured at
    the time of bounce. The rest of the columns are: final and
    explosion times in seconds after bounce, the 
    mass inside the shock wave at the onset of explosion
    ($M_{\mathrm{sh,e}}$), the kind of explosion (neutrino driven, $\nu$; magnetorotational, M),
    the final explosion energy, $E_{\mathrm{exp}}$, the free energy
    of the rotation of the PNS, $\Frotpns$, and the rotational energy of the PNS, $\Erotpns$, at the end of the simulation,
    the type of
    remnant (the acronym BH is set in boldface if BH formation occurs
    during the simulation; in this case, the final simulation time is
    the BH formation time) and its mass at the final time,
    $M_{\mathrm{r}}$, and an estimate for the expelled mass of Ni,
    $M_{\mathrm{Ni}}$.
  }
  \label{Tab:Globerview}
\end{table*}

\begin{figure}
  \centering
  \includegraphics[width=\linewidth]{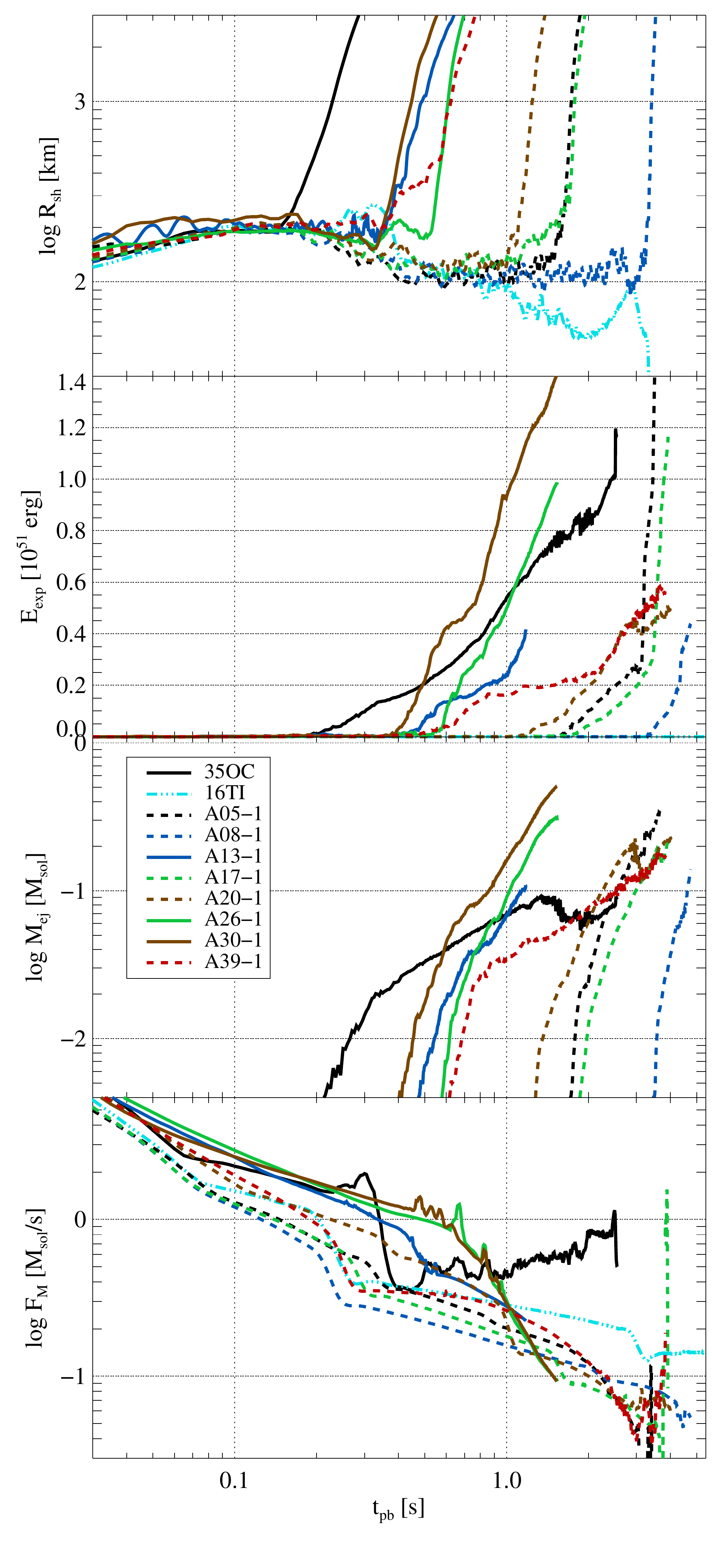}
  \caption{
    Top to bottom: 
    shock radius, explosion energy, ejecta mass, and mass flux through
    the shock wave as functions of 
    post-bounce , $t_{\rm pb}$.
  }
  \label{Fig:globvars}
\end{figure}

\begin{figure}
  \centering
  \includegraphics[width=\linewidth]{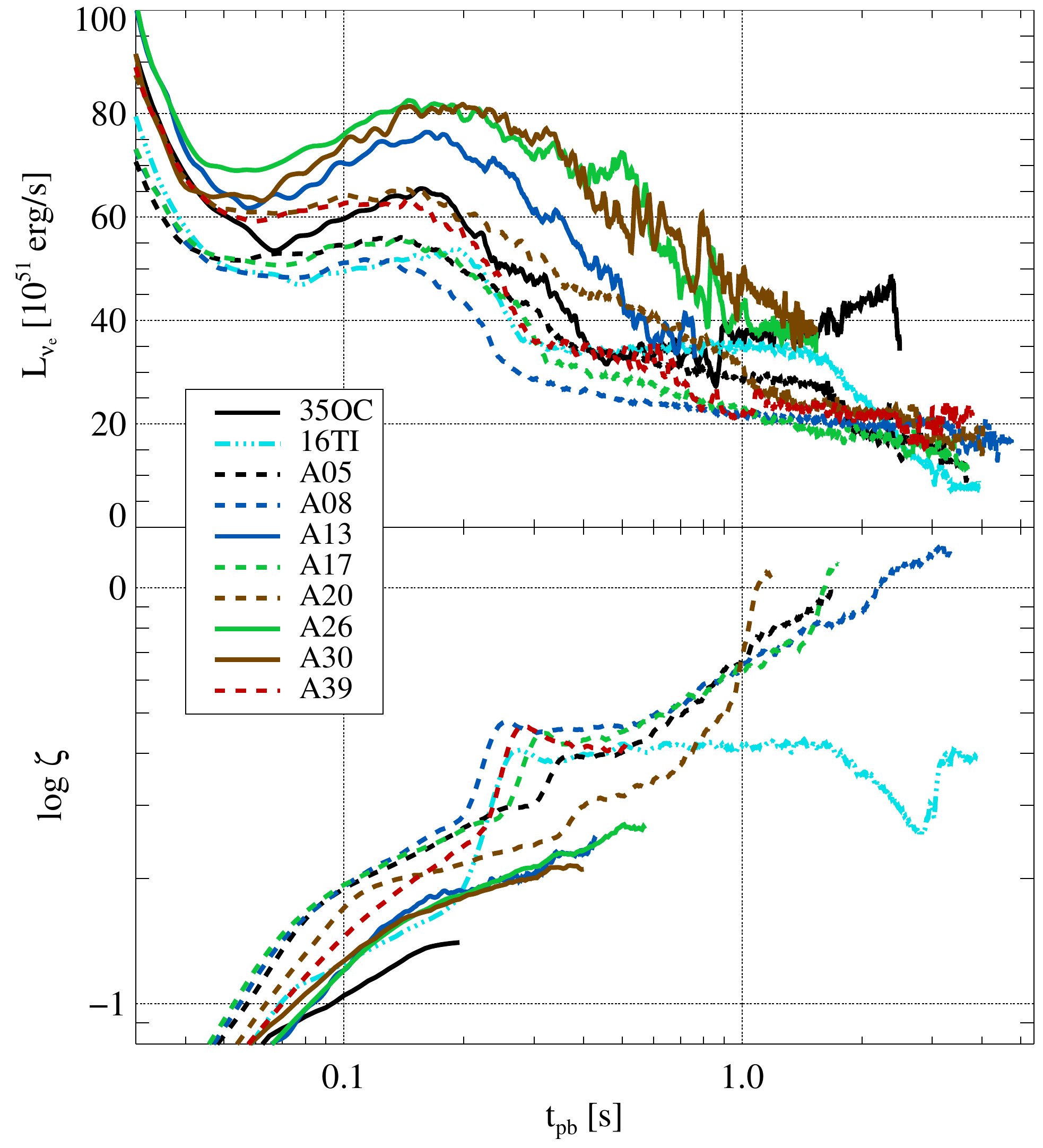}
  \caption{
    Top panel: luminosities of electron neutrinos.
    Bottom panel: ratio of the total neutrino luminosity to the mass accretion rate
    through the shock, $\zeta$.
  }
  \label{Fig:globvars-lumg}
\end{figure}

\begin{figure}
  \centering
  \includegraphics[width=\linewidth]{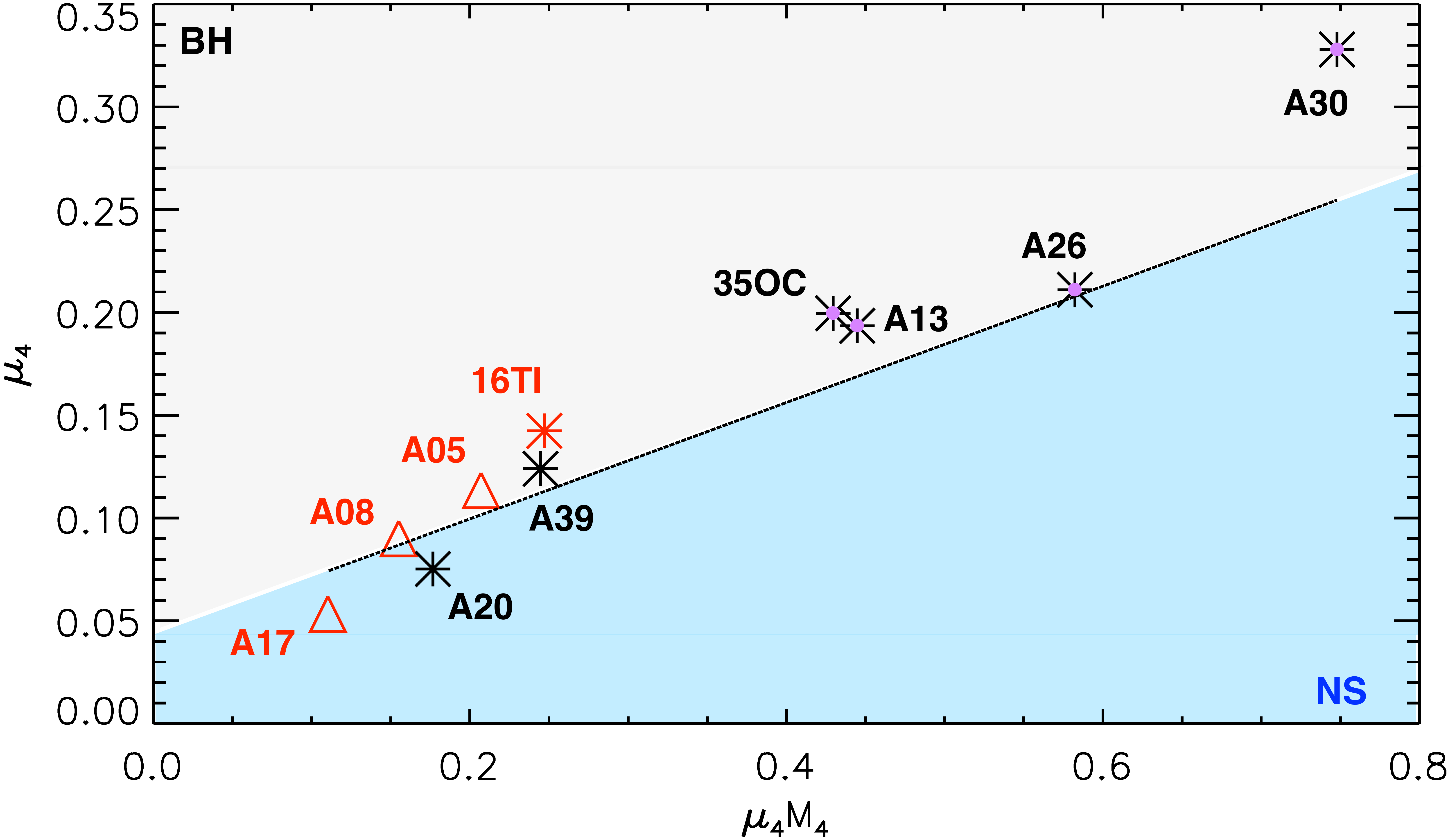}
  \caption{Representation of the computed models in the
    $\mu_4M_4-\mu_4$ plane. The black line corresponds
    to $y=0.283x+0.043$, where $y=\mu_4$ and $x=\mu_4 M_4$.  Black
    asterisks identify models of \tabref{Tab:Globerview} that yield
    BHs, while red triangles correspond to models that do not form
    BHs. The red asterisk represents \modl{16TI}, which has not formed
    a BH by the end of the computed time. Asterisks with an
      overlaid magenta circle correspond to models that develop early
      magneto-rotational explosions.}
  \label{Fig:mu4M4}
\end{figure}

\begin{figure}
  \centering
  \includegraphics[width=\linewidth]{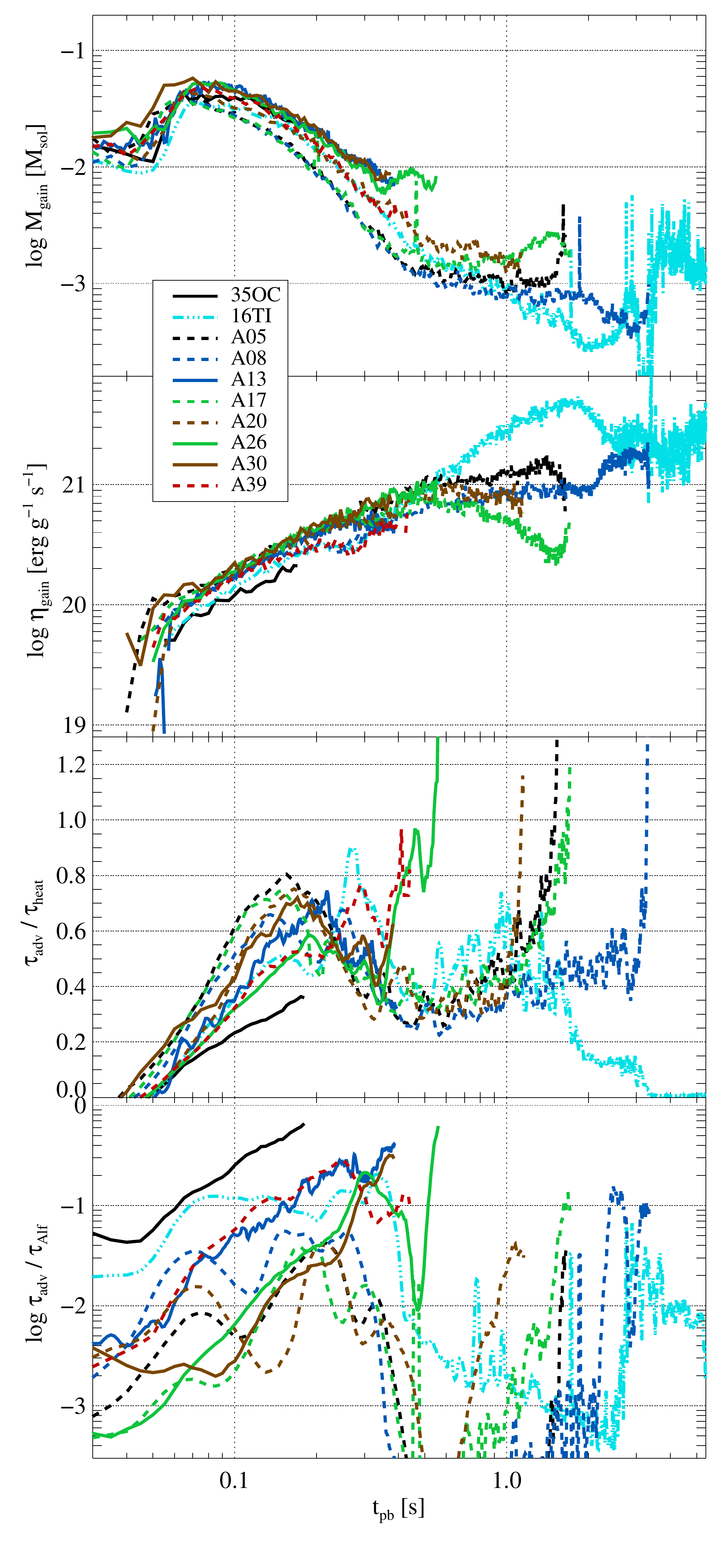}
  \caption{
    Top to bottom: mass in the gain layer, heating efficiency, ratio
    between advection and heating timescales, and between the
    advection and the \Alfven timescales.  The lines and their colors are as in Fig.\,\ref{Fig:globvars-lumg}. The evolution is only displayed until the definitive shock expansion sets in.
  }
  \label{Fig:globvars-gain}
\end{figure}

We summarize the outcomes of our simulations in
\tabref{Tab:Globerview}, listing the explosion mechanism and the kind
of compact remnant in comparison to the predictions of \AguiDet for
the evolution of the stars where applicable.  \Modl{16TI} fails to
launch an explosion during the entire time simulated and will produce
a BH within a few more seconds.  All other models experience delayed
shock revival initiated by a combination of neutrino heating,
non-spherical fluid flows, and magnetorotational stresses.  The
importance of these effects as well as the time of explosion vary
greatly across the mass range.  In some cases, the explosion is
sufficiently strong to quench accretion onto the PNS and prevent BH
formation, but most cores, in particular at higher stellar masses
($M_{\textsc{ZAMS}}\ge 20M_\odot$), are likely to end up as BHs.

In the following, we will describe the most important features
characterizing the evolution of several groups of models:
\begin{enumerate}
\item neutrino-driven explosions leading to NS or BH formation: \modls{A05,
    A08, A17, A20, A39};
\item MHD explosions with BH formation: \modls{A13, A26, A30, 35OC};
\item the failed explosion of \modl{16TI}.
\end{enumerate}

\subsection{Neutrino-driven explosions}
\label{sSek:NuDrEx}

We will focus our discussion on models whose explosion is chiefly
driven by neutrinos. Neutrino-driven explosions require (at least)
that the heating time scale due to neutrinos be comparable or larger
than the advection timescale through the gain layer.  \Modls{A05},
\modelname{A08}, \modelname{A17}, and \modelname{A20} undergo shock
revival with a delay after core bounce between $t_{\mathrm{exp}}
\approx 1.0 \, \sek$ (\modelname{A20}) and $t_{\mathrm{exp}} \approx
3.2 \, \sek$ (\modelname{A08}) (see evolution of the maximum shock
radius in \panel{top panel} of \figref{Fig:globvars}).  \Modl{A39}
explodes much earlier ($t_{\mathrm{exp}} \approx 0.5 \, \sek$) and,
showing a more pronounced influence of the magnetic field, could be
considered an intermediate case between this class and that of MHD
explosions (Sec.\,\ref{sSek:MRexp}).

After the onset of the explosion, the shock wave expands at high
speeds and the diagnostic explosion energies and ejecta masses
(\figref{Fig:globvars} \panel{mid and bottom panels}) rise.  By the
end of the simulations, the explosion energies are around $\eej
\approx \zehnh{5}{50} \, \ergs$ carried by $\mej \approx 0.1...0.4 \,
\msol$ of unbound gas.  Both quantities keep increasing throughout the
final phases of the simulation.  \Modl{A17} is an exception in that it
shows a very rapid rise of $\eej$ by almost $10^{51} \, \erg$ within
about $0.5 \, \sek$ (starting at a post-bounce time $t_{\rm pb}\approx
3.1\,\sek$) without a similar growth of $\mej$.  This quick rise,
which does not show any sign of levelling off, corresponds to the
activation of a wind powered by the magnetic spin-down of the PNS,
which by then has turned into a PM due to a growth of the surface
field strength.  This late-time evolution does not interfere with the
revival of the SN shock wave.

Due to the large rotational energy existing in the stellar progenitors
of all the models in this study, the compact remnant left by their
core collapse stores a rather large rotational energy, $\Erotpns$
(Tab.\,\ref{Tab:Globerview}). Thus, even neutrino-driven explosion
models harbor an additional source of energy that can be transferred
to the ejecta. It is, nevertheless, arguable whether this extra energy
will be a sizeable fraction of $\Erotpns$ or only the (significantly
smaller) free rotational energy of the PNS, $\Frotpns$ (defined in
Eq.\,\eqref{eq:freerotenergy} below).

The models of this group tend to possess the least compact cores among
the progenitors considered here.  Their density profiles, $\rho(m)$
(\figref{Fig:initmodels} \panel{top panel}; dashed lines), exhibit a
strong decrease at mass coordinates $m < 2 \, \msol$ near the location
of the surface of the inner core, which show up as discontinuities in
the specific entropy and the electron fraction
(\figref{Fig:initmodels} \panel{middle panel}).  The comparably low
densities further out correspond to moderate values of the compactness
parameter, $\xi_{2.5}$, computed at $m=2.5M_\odot$ according to
Eq.\,\eqref{eq:xi}, are listed in \tabref{Tab:Globerview} (see also
\figref{Fig:initmodels}, \panel{bottom panel}). These values are
significantly below the threshold $\xi_{2.5}\approx 0.45$ of
explodability
\citep{OConnor_Ott__2011__apj__BlackHoleFormationinFailingCore-CollapseSupernovae}
and, hence, these stellar models should easily yield successful
supernova explosions.  During the pre-explosion phase, these models
develop low mass accretion rates through the stalled shock due to the
low densities beyond the densest central part of the star
(\figref{Fig:globvars}, \panel{top
  panel}). \cite{Ertl_et_al__2016__apj__ATwo-parameterCriterionforClassifyingtheExplodabilityofMassiveStarsbytheNeutrino-drivenMechanism}
defined a two-parameter criterion for classifying the explodability of
massive stars by the neutrino-driven mechanism of
\textit{non-rotating} models. These two parameters are $M_4$, defining
the normalized enclosed mass for a dimensionless entropy per nucleon
of $s=4$,
\begin{align}
M_4 \equiv m(s=4) / M_\odot
\label{eq:M4}
\end{align}
and $\mu_4$, the normalized mass derivative at this location
\begin{align}
\mu_4 \equiv \left | (dm/M_\odot)/ (dr/1000\,\text{km})\right |_{s=4}
\label{eq:mu4}
\end{align}
According to
\cite{Ertl_et_al__2016__apj__ATwo-parameterCriterionforClassifyingtheExplodabilityofMassiveStarsbytheNeutrino-drivenMechanism},
the parameters $\mu_4$ and $\mu_4M_4$ tightly correlate with the
neutrino luminosity and the mass accretion rate through the stalled
shock, respectively. Our neutrino-driven explosion models, possess the
smallest values of $M_4$, of $\mu_4$, and of $\mu_4 M_4$
of all the sample (\tabref{Tab:Globerview}). Although our models are
fast rotators, we also notice that small values of $\mu_4$ correlate
with the small neutrino luminosities (\figref{Fig:globvars-lumg},
\panel{upper panel}). Likewise, the small values of the product $\mu_4
M_4$, correlate with the mass accretion rate onto the PNS (or through
the shock; \figref{Fig:globvars} \panel{bottom panel}).
Figure\,\ref{Fig:mu4M4} represents the models of
\tabref{Tab:Globerview} in the $\mu_4M_4-\mu_4$ plane. The black line
represents an approximate division of the plane in BH forming and NS
forming models according to
\cite{Ertl_et_al__2016__apj__ATwo-parameterCriterionforClassifyingtheExplodabilityofMassiveStarsbytheNeutrino-drivenMechanism}.
Specifically, the line that we represent is $y=k_1 x +k_2$, where
$y=\mu_4N_4$, $x=\mu_4$, and the coefficients $k_1=0.283$ and
$k_2=0.043$ are taken from one of the sets of fits of
\cite{Ertl_et_al__2016__apj__ATwo-parameterCriterionforClassifyingtheExplodabilityofMassiveStarsbytheNeutrino-drivenMechanism}.
We observe that not only BH forming models lie above the division
line, but also some models, and may hardly form an NS in a longer
timescale than spanned by our calculations. Furthermore, below the
division line there are models which will likely form NS later on
\citep[as expected according to the classification
of][]{Ertl_et_al__2016__apj__ATwo-parameterCriterionforClassifyingtheExplodabilityofMassiveStarsbytheNeutrino-drivenMechanism}
as well as a BH forming case, precisely that of \modl{A20}, which
produces a neutrino-driven explosion. Using other values for $k_1$ and
$k_2$, the two-parameter classification criterion does not improve
much \citep[we have used all the possible combinations of $k_1$ and
$k_2$ from Table 2
of][]{Ertl_et_al__2016__apj__ATwo-parameterCriterionforClassifyingtheExplodabilityofMassiveStarsbytheNeutrino-drivenMechanism}.
For the set of models at hand, values $k_1\sim 0$, $k_2\sim 0.12$
yield a better classification between BH forming and NS forming
models.Nevertheless, the parameters $\xi_{2.5}$, $M_4$, and $\mu_4$ do
not gather direct information about the rotational and magnetic
properties of the stellar progenitor. Hence, they (or their
combinations) may only be approximately used to predict whether an
explosion will be neutrino-driven or magneto-rotational. We have
explored a number of additional indicators for that end.

First, we have searched for the existence of simple correlations
between the rotational energy of the progenitor, the explosion type,
and the compact remnant. Precisely, we compute the rotational energy
inside $m=M_4$, measured in units of $\zehn{48}\,\erg$,
$\mathcal{T}_4$.  This quantity does not discriminate between
exploding or non-exploding models. For instance, \modl{16TI}, which
fails to explode, has a similar value of $\mathcal{T}_4$ as
\modl{A26}, which explodes magneto-rotationally. There is, however, a
broad correlation between the explosion type and the value of
$\mathcal{T}_4$. Larger values of $\mathcal{T}_4\gtrsim 2.1$ yield
magneto-rotational or mixed type explosions (where magneto-rotational
effects act together with neutrinos). Conversely, values
$\mathcal{T}_4\lesssim 2.1$ associate with neutrino-driven
explosions. The compact remnant type also correlates with
$\mathcal{T}_4$.  Small values of $\mathcal{T}_4 (\lesssim 1.4)$ yield
PNSs as compact remnants, while large values of $\mathcal{T}_4(\gtrsim
1.4)$ produce BHs (Tab.\,\ref{Tab:Globerview}). We add a cautionary
note here. The final fate of the compact remnant is (to some extent)
sensitive to the dimensionality of the model. Models in 3D display a
significantly smaller mass growth rate of the PNS after $t_{\rm
  pb}\sim
0.2\,\sek$\citepalias{Aloy__2021__MonthlyNoticesoftheRoyalAstronomicalSociety__MagnetorotationalCoreCollapseofPossibleGRBProgenitorsII.FormationofProtomagnetarsandCollapsars,Obergaulinger__2021__mnras__MagnetorotationalCoreCollapseofPossibleGRBProgenitorsIII.ThreeDimensionalModels}. Hence,
BH formation (if at all happens) may take some more time in 3D than in
2D.

Second, we stress the fact that the structure of the star is
  formed by a set of magnetized layers intertwined with non-magnetized
  ones. We may compute the average coherence length of the poloidal
  magnetic field within the inner $2.5M_\odot$ as
\begin{align*}
\overline{l}_{\rm pol}\equiv \frac{1}{N_{\rm layers}}\sum_{i=1}^{N_{\rm layers}} \Delta_i ,
\end{align*}
where $\Delta_i$ is the radial width of a magnetized layer and $N_{\rm
  layers}$ is the number of magnetized layers up to a mass coordinate
$m=2.5M_\odot$. The choice of the mass coordinate $m=2.5M_\odot$ is
arbitrary, but we have checked that the results that follow do not
sensitively depend on the exact mass considered. Since it is customary
to evaluate the compactness ratio at $m=2.5M_\odot$, we have adopted
this value. We note that, e.g., in most cases $M_4 \lesssim
2.5M_\odot$. Evaluation of the coherence length until $m=M_4$ yields
qualitatively similar results.
 
If we denote $r_{2.5}$ the radius corresponding to
  $m=2.5M_\odot$, we can build the following dimensionless ratio
\begin{align}
\lambda_{2.5}\equiv \overline{l}_{\rm pol} / r_{2.5}
\label{eq:lambda25}
\end{align}
Values of $\lambda_{2.5}$ close to 1 mean that the inner core of the
star is all magnetized and that the magnetic field connects stellar
matter from the centre to $r_{2.5}$. In
\citetalias{Aloy__2021__MonthlyNoticesoftheRoyalAstronomicalSociety__MagnetorotationalCoreCollapseofPossibleGRBProgenitorsII.FormationofProtomagnetarsandCollapsars}
we stressed the fact that a larger poloidal magnetic field coherence
length helps develop magnetorotational explosions. This is because
after bounce, the PNS is magnetically connected to the gain region and
may transfer angular momentum and energy there by magneto-centrifugal
effects. In Tab.\,\ref{Tab:Globerview} we observe the clear
correlation between $\lambda_{2.5}$ and the kind of explosion. If
$\lambda_{2.5}\gtrsim 0.3$, magnetic effects are important for the
explosion success, while if $\lambda_{2.5}<0.1$ neutrinos are the main
driver for the explosion. It is significant that the magnetic field
strength within the inner region of the progenitor star does not
clearly correlate shows some correlation with the explosion
mechanism. We support this statement computing the volume-averaged
poloidal (toroidal) magnetic field strength in the region interior to
$m=2.5M_\odot$, $\overline{B}^{\,\rm p}_{\rm 2.5}$
($\overline{B}^{\,\rm t}_{\rm 2.5}$; see Tab.\,\ref{Tab:Globerview}).
Models with $\overline{B}^{\,\rm p}_{\rm 2.5}\lesssim
\zehnh{1.3}{10}\,$G develop neutrino-driven explosions. However,
possessing a large volume-averaged magnetic field strength does not
guarantee that neutrinos do not play any role in the explosion. This
is the case, e.g., in \modl{A39}, which yields a mixed-type explosion
although it has a volume-averaged magnetic field very similar to that
of \modl{A17}, which is one of the clearest examples of
magneto-rotationally driven-explosion. CHE models show poloidal and
toroidal fields $\sim 1-2$ orders of magnitude smaller than non-CHE
models.  This fact does not prevent that some CHE models develop
magneto-rotational explosions. As in the case of the quantity
$\mathcal{T}_4$, $\lambda_{2.5}$ does not allow predicting whether a
specific model may or may not explode.

\begin{figure*}
  \centering
  \includegraphics[width=\linewidth]{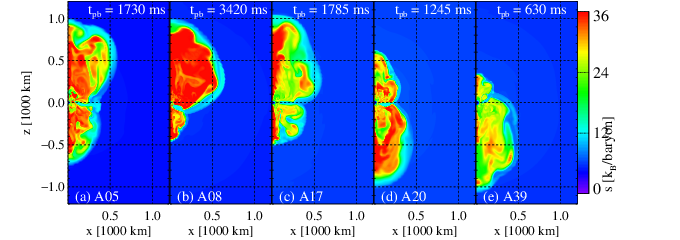}
  \includegraphics[width=\linewidth]{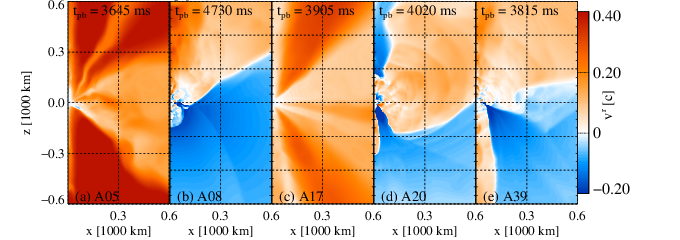}
  \caption{
    Top: Entropy maps of neutrino-driven explosions at the time when the
    shock wave has reached a maximum radius of $r_{\mathrm{sh}}
    \approx 1000 \, \km$.
    Bottom: maps of radial velocity towards the end of the runs.
  }
  \label{Fig:2dexpnud}
\end{figure*}

Despite the favourable conditions for explosions, it takes more than
one second to achieve shock revival.  Relative to the models exploding
earlier due to MHD effects, these cores tend to emit neutrinos at
lower total luminosities (see \figref{Fig:globvars-lumg}, \panel{upper
  panel}).  The luminosities are not only lower in absolute values,
but also relative to the gravitational energy of the PNSs.  While, the
production of neutrinos seems less efficient in this group of models,
their effect on the dynamics is higher.  We can quantify this effect
with the ratio of the total neutrino luminosity to the mass accretion
rate through the shock, $\zeta = L_{\nu} / (\dot{M} c^2)$
(\panel{bottom panel}).  Throughout the entire runs, \modls{A05},
\modelname{A08}, \modelname{A17}, \modelname{A20}, and \modelname{A39}
show higher values of $\zeta$ than models exploding
magnetorotationally.  Their shock revival sets in when at $\zeta
\gtrsim 1$.  At that time, the mass in the gain layer has decreased to
a few $10^{-3} \, \msol$ (\figref{Fig:globvars-gain}, \panel{top
  panel}), while the heating efficiency, $\eta_{\mathrm{gain}} =
Q_{\nu, \mathrm{gain}} / M_{\mathrm{gain}}$, rises until the explosion
sets in (\figref{Fig:globvars-gain}, \panel{second panel}). The
neutrino heating timescale in the gain layer, $\tauhtg$, is
considerably longer than the advection timescale, $\tauadv$, except
for an episode around $\tpb \approx 0.1...0.2 \, \sek$.  The average
of the ratio $\tautau$ over the gain layer(\figref{Fig:globvars-gain}
\panel{third panel}) rapidly rises above unity, but only immediately
before the shock wave revives.  We note that, compared to \modl{35OC},
all new models come closer to fulfilling the explosion criterion of
$\tautau = 1$ during an earlier phase of their evolution ($\tpb =
0.1...0.2 \, \sek$).  The heating rate, however, does not suffice to
increase the total energy of the gain layer to positive values.  The
failure can in part be attributed to a relatively low activity of
non-spherical flows during this time.  Furthermore, the magnetic
energies in the gain layer are low and correspond to ratios between
the advection time and the \Alfven travel time across the gain layer,
$\tautaum$, that are on average much less than unity (\panel{bottom
  panel}).  Hence, even taking into account a latitudinal variation of
$\tautaum$, the magnetic field contributes very little to reviving the
shock wave.

After this phase, a reduction of the mass accretion rate and a
moderate contraction of the shock wave characterize the models, which
settle into a state maintained for a time between several hundreds of
milliseconds and about $3 \, \sek$.  Fairly constant neutrino
luminosities translate into a timescale ratio $\tautau \sim
0.3...0.6$.  In \modls{A08}, \modelname{A17}, and \modelname{A20},
shock runaway is triggered by the quick decrease of the ram pressure
corresponding to the infall of a shell interface located at a mass
coordinate $m = 2.1...2.4 \, \msol$, while \modelname{A05} reaches the
transition to explosion without such an additional impulse due to a
slow reduction of $\tauhtg$ over the course of about half a second.
In the case of \modl{A39}, a non-negligible magnetic energy in the
gain layer adds to the effect of neutrino heating and allows for an
explosion at a much earlier time than in the other models
($t_{\mathrm{exp}} \sim 0.4 \, \sek$).

\begin{figure*}
  \centering
  \includegraphics[width=0.48\textwidth]{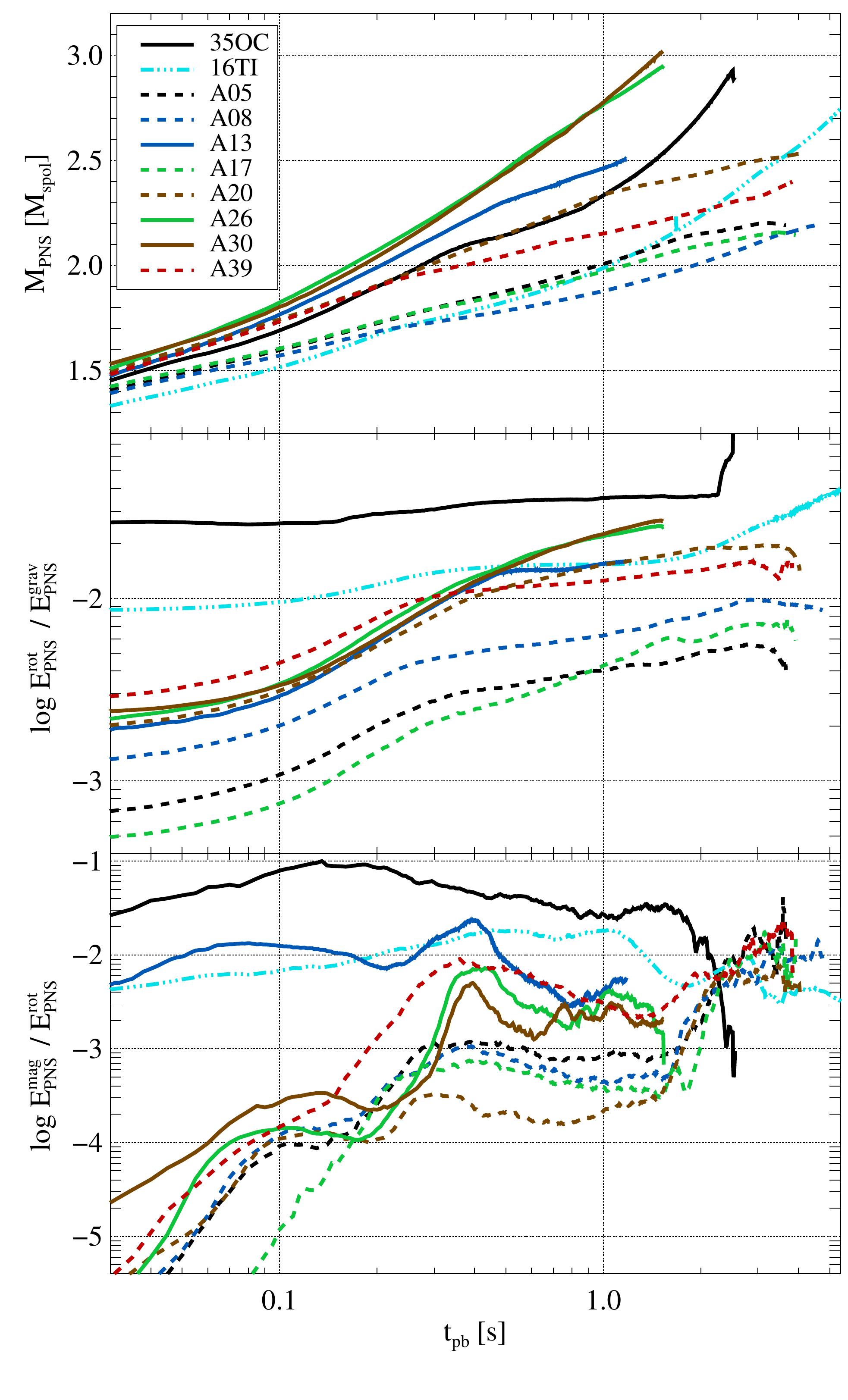}
  \includegraphics[width=0.48\textwidth]{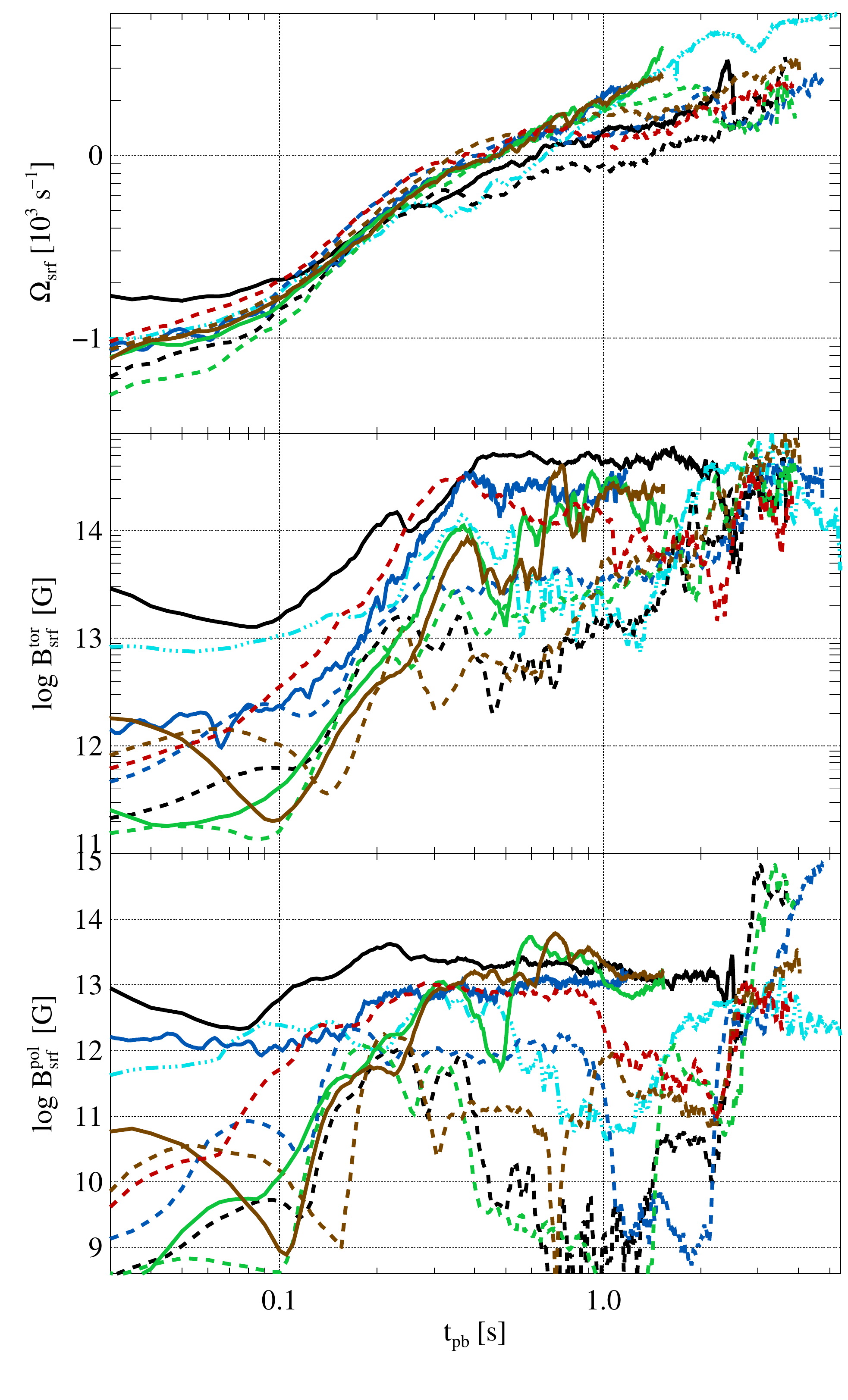}
  \caption{
    Left panel, top to bottom: evolution of the mass of the PNSs, the
    rotational energy normalized by the gravitational energy, and
    the magnetic energy normalized by the rotational energy.
    Right panel, top to bottom: averages of the rotational
    velocity, the toroidal and poloidal magnetic field strengths over
    the PNS surfaces.
  }
  \label{Fig:globvars-2}
\end{figure*}

The constraint of axisymmetry causes the ejecta to expand along the
poles (\figref{Fig:2dexpnud}).  Variations in the fluxes of mass and
energy lead to asymmetries of the outflows ranging from a moderate
north-south imbalance to a unipolar geometry.  Typically, the outflows
coexist with cold inflows at lower latitudes that feed the growth of
the PNS throughout the entire evolution after shock revival.  The
\panel{bottom panel} of \figref{Fig:2dexpnud} displays the structure
of the inflows/outflows in the central 600 km of the five
neutrino-driven explosion models shortly before the end of the
simulations. The inflows can take the form of narrow streams at high
latitudes (\modelname{A20}), at lower latitudes (\modelname{A39}), as
well as one-sided inflows across an entire hemisphere
(\modelname{A08}).  Transitions between these states also occur.
Consequently, the PNS mass (\figref{Fig:globvars-2}, \panel{top
  panel}) exceeds $\MPNS = 2 \, \msol$ before the end of the
simulations in all cases.  For \modls{A20} and \modelname{A39}, the
final values of $\MPNS \approx 2.55 \, \Msol$ and $\MPNS \approx 2.42
\, \Msol$ correspond to BH formation.  While \modl{A08} was terminated
with a considerably lower PNS mass of $\MPNS \approx 2.2 \, \msol$,
the rate of accretion is similar to that of the aforementioned models
and, thus, the collapse to a BH is a plausible outcome, too.  Note,
however, that the long-term evolution of the PNS is not necessarily
monotonic in terms of, e.g. its mass growth, especially if
sufficiently strong poloidal magnetic fields develop
\citepalias{Aloy__2021__MonthlyNoticesoftheRoyalAstronomicalSociety__MagnetorotationalCoreCollapseofPossibleGRBProgenitorsII.FormationofProtomagnetarsandCollapsars},
as it is the case of \modl{A08}.

\Modls{A05} and \modelname{A17} show a transition to a phase in which
accretion ceases to occur after $t_\mathrm{pb}>3\,\sek$.  No inflows
reach the PNS.  While matter is ejected at all latitudes, the
velocities and mass fluxes of the outflows are highest inside wide
cones around the rotational axis (\figref{Fig:2dexpnud} \panel{bottom
  panel}).  At radii of a few 1000 km, these outflows are directed
into wide polar jets (\figref{Fig:2dexpnud}, \panel{upper panel}).
This configuration causes the PNSs to lose mass at rates of the order
of $\dot{M}_{\mathrm{PNS}} \sim - 10^{-2} \, \msol \, \sek^{-1}$ and
signal a transient episode of PM spin-down (transient episodes of PM
spin-down were also observed in models with sufficiently strong
poloidal field in
\citetalias{Aloy__2021__MonthlyNoticesoftheRoyalAstronomicalSociety__MagnetorotationalCoreCollapseofPossibleGRBProgenitorsII.FormationofProtomagnetarsandCollapsars}).
We cannot precisely compute from our models the additional energy that
the PM spin-down could add to the explosion. However, it is at least
of the order of the free energy of the PNS differential rotation
(e.g.,\cite{Dessart_et_al__2012__apj__TheArduousJourneytoBlackHoleFormationinPotentialGamma-RayBurstProgenitors};
\citetalias{Aloy__2021__MonthlyNoticesoftheRoyalAstronomicalSociety__MagnetorotationalCoreCollapseofPossibleGRBProgenitorsII.FormationofProtomagnetarsandCollapsars})
or, at most, of the order of the PNS rotational energy
\citep[e.g.][]{Metzger_et_al__2011__mnras__Theprotomagnetarmodelforgamma-raybursts,
  Metzger_et_al__2015__mnras__Thediversityoftransientsfrommagnetarbirthincorecollapsesupernovae}. Any
process capable of redistributing angular momentum (e.g., the
magnetorotational instability, viscosity, etc.) may potentially tap
the shear energy and contribute to the SN explosion. For instance,
viscosity may add extra heating to the postshock region
\citep{Thompson_Quataert_Burrows__2004__ApJ__Vis_Rot_SN}, while the
magnetorotational instability may lead to a further growth of the
magnetic field, driving a magnetorotational explosion
\citepalias{Obergaulinger_Aloy__2020__mnras__MagnetorotationalCoreCollapseofPossibleGRBProgenitorsIExplosionMechanisms}. As
in
\cite{Dessart_et_al__2012__apj__TheArduousJourneytoBlackHoleFormationinPotentialGamma-RayBurstProgenitors}
and
\citetalias{Aloy__2021__MonthlyNoticesoftheRoyalAstronomicalSociety__MagnetorotationalCoreCollapseofPossibleGRBProgenitorsII.FormationofProtomagnetarsandCollapsars}
we compute the free rotational energy by subtracting the rotational
energy of a rigidly rotating body with the same angular momentum,
$J_\ppns$ and inertial momentum, $I_\ppns$, from the total rotational
energy of the PNS,
\begin{align}
\Frotpns = \Erotpns - \frac{J_\ppns^2}{2
  I_\ppns}
  \label{eq:freerotenergy}
\end{align}
(see \tabref{Tab:Globerview}).  In the case of \modl{A17}, this value
is greater than the explosion energy at the end of the simulation,
$\Frotpns \approx 2.2 \times 10^{51} \,\erg$, while for
\modelname{A05} it accounts only for a minor correction to the
explosion.

All PNSs possess a highrotational energy $\Erotpns > 10^{52}\,\erg$
(see \tabref{Tab:Globerview}).  Final values of the rotational energy
correspond to ratios $- \Erotpns / E^{\mathrm{grav}}_\ppns \sim
0.5...2 \times 10^{-2}$ (\figref{Fig:globvars-2}, \panel{second
  panel}), i.e., a range in which the PNS might be susceptible to
non-axisymmetric instabilities.  Besides amplification processes
inside the PNS, the magnetization of the accreted matter drives the
evolution of the magnetic energy.  The final magnetic energies are
typically two orders of magnitude lower than the rotational energies
(\figref{Fig:globvars-2}, \panel{bottom left panel}).  Typical surface
rotation rates reach a few $1000 \, \isek$ (\panel{upper right panel})
with models exploding mostly due to neutrino heating found at the
lower end of the distribution of $\Omega_{\mathrm{srf}}$ (hereafter,
the subscript ``srf'' denotes average quantities over the PNS
surface).  The average magnetic fields on the PNS surfaces are
dominated by a toroidal component of the order of
$B^{\mathrm{tor}}_{\mathrm{srf}} \sim 10^{14\ldots 15} \, \Gauss$
(\figref{Fig:globvars-2}, \panel{mid right panel}).  The poloidal
components (\panel{bottom right panel}) tend to be considerably weaker
during most phases of the evolution, in some cases down to a mere
$B^{\mathrm{pol}}_{\mathrm{srf}} \sim 10^{9} \, \Gauss$.  As stated
above, the suppression of the poloidal field proceeds in concert with
the burial of the surface field by non-magnetized accreted
layers. Likewise, the accretion of magnetized layers can lead to a
quick rise of the poloidal field strength, exceeding
$B^{\mathrm{pol}}_{\mathrm{srf}} \sim 10^{14} \, \Gauss$ and reaching
equipartition with the toroidal component.  Hence, the surface
poloidal magnetic field of the models with smaller mass among the
group that yields neutrino-driven explosions (\modelname{A05},
\modelname{A08}, and \modelname{A17}) grows to magnetar field strength
after, at least $t_\mathrm{pb}\sim 2.9\,\sek$.  Such conditions can
favour the spin-down of the PNS and the transfer of its shear or
rotational energy to the surrounding gas, which then is accelerated
and forms the aforementioned wind-like outflows of \modls{A05} and
\modelname{A17}.

\subsection{Magnetorotational explosions}
\label{sSek:MRexp}

Models whose explosions are launched mainly by magnetic fields and
rotation (\modelname{35OC}, \modelname{A13}, \modelname{A30}, and, to
a slightly lower degree, \modelname{A26}) exhibit shock revival after
at most $t_{\mathrm{exp}} \lesssim 0.5 \, \sek$
(\figref{Fig:globvars}) and a steady increase of the explosion
energies and masses. For the two models based on progenitors with
higher masses, \modelname{A26} and \modelname{A30}, the energy
increase proceeds faster than for the lower-mass \modl{A13} as well as
for \modl{35OC}.  The progenitors of \AguiDet end with $E_\mathrm{exp}
\gtrsim 0.8 \times 10^{51} \, \erg$ and $\mej \gtrsim 0.3 \, \Msol$.

The progenitors of this group possess more compact cores than those
producing neutrino-driven explosions, with higher densities and
compactness parameters outside a mass coordinate of $m \approx 1.8 \,
\msol$ (\figref{Fig:initmodels}).  Models driving magnetorotational
explosions also have larger values of $M_4$ and of $\mu_4$ as their
neutrino-driven counterparts.  These differences, related to their Fe
cores containing more mass, translate into higher mass accretion rates
through the shock (\panel{bottom panel} of \figref{Fig:globvars}) and
lower neutrino luminosities \wrt the mass accretion rate, with $\zeta$
consistently remaining below those of neutrino-driven explosions
(\figref{Fig:globvars-lumg} \panel{bottom panel}).  These conditions
make neutrino-driven shock revival more difficult to achieve.

The mass contained in the gain layers of these four models is on the
upper end of the distribution (\panel{top panel} of
\figref{Fig:globvars-gain}), while the rotational energies of the gain
layer are lower than in the case of neutrino-driven explosions.  The
non-radial kinetic energies, on the other hand, do not systematically
deviate from the other models, suggesting a similar level of
hydrodynamic instabilities and turbulence.  At least within the group
of models from \AguiDet, the same holds for the heating efficiency,
$\eta_{\mathrm{gain}}$ (\figref{Fig:globvars-gain} \panel{mid panel}).
As a consequence, the models show a similar evolution of the
advection-to-heating timescale ratio, $\tautau$ with an early rise
that by $\tpb \lesssim 200 \, \ms$ stops around $\tautau \approx 0.8$,
i.e., at values that can be regarded insufficient for launching an
explosion (\panel{bottom panel}).

\Modl{35OC} possesses a relatively high magnetic energy in the gain
layer already shortly after bounce.  The three models from \AguiDet
start with a much weaker magnetization in the gain layer, but the
accretion of extended layers with non-zero magnetic field
(cf.~\figref{Fig:init-angmom}) causes the magnetic energy to rise to
values around $E^{\mathrm{mag}}_{\igain} \gtrsim 10^{47} \, \erg$.
\Modls{A13} and \modelname{A30} explode shortly after that point.  In
these cases, the average ratio between the advection and the \Alfven
timescale, $\tautaum$, is still significantly lower than unity
(\figref{Fig:globvars-gain}, \panel{bottom panel}).  However, since
the magnetic field is much stronger along the rotational axis than at
lower latitudes, the \Alfven timescale becomes comparable to the
advection timescale locally, giving rise to polar shock revival
\citepalias[as discussed
in][]{Obergaulinger_Aloy__2020__mnras__MagnetorotationalCoreCollapseofPossibleGRBProgenitorsIExplosionMechanisms}.
The shock wave of \modelname{A26} ceases to propagate after a short
outward displacement and is revived only after a delay of another
about $150 \, \ms$, when the timescales for neutrino heating and for
\Alfven waves locally reach values similar to that of advection.

\begin{figure}
  \centering
  \includegraphics[width=\linewidth]{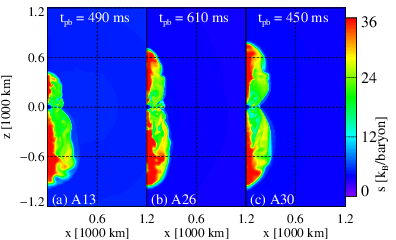}
  \includegraphics[width=\linewidth]{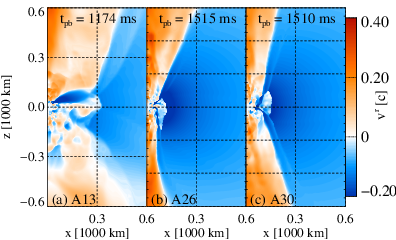}
  \caption{
    Same as \figref{Fig:2dexpnud}, but for magnetorotational
    explosions.
  }
  \label{Fig:2dMHDexp}
\end{figure}

To a larger degree than for neutrino-driven shock revival, the
different ejecta, whose entropy distributions are displayed in the
\panel{top panel} of \figref{Fig:2dMHDexp} at the time when the
maximum shock radius reaches $1000 \, \km$, are propagating in narrow
jets along the rotational axis.  The flow speeds in the jets reach up
to $v^r \gtrsim c/3$ at radii of a few 1000 km with models
\modelname{A26} and \modelname{A30} producing faster ejecta, followed
by \modl{A13}.  Inflows near the equatorial plane (see
\panel{bottom panel} of \figref{Fig:2dMHDexp}), found in all models,
are persistent and lead to a comparably rapid growth of the PNS mass.
\Modls{A26} and \modelname{A30} surpass the limit for stability of
non-rotating neutron stars within less than $\approx 0.6 \, \sek$.
The growth proceeds until $\MPNSm{A26,A30} \approx 3 \, \Msol$, at
which point not even the rotational energy corresponding to
$\Erotpns^{\modelname{A26},\modelname{A30}}\approx 0.03 \,
E^{\textrm{grav},\modelname{A26,A30}}_{\ppns}$ and a high degree of
differential rotation can prevent the PNS from collapsing to a BH.
Reaching gravitational instability takes longer for the less compact
\modl{A13}.  The magnetic influence on the explosion is not reflected
in a particularly high magnetic energy of the PNSs.  Compared to
\modl{35OC}, the ratio of magnetic to rotational energy is lower by at
least an order of magnitude.  Indeed, the evolution of the ratio
$E^\textrm{mag}_\ppns/\Erotpns$ is very different in CHE models
compared to the models of
\cite{Woosley_Heger__2006__apj__TheProgenitorStarsofGamma-RayBursts}. The
former models possess much larger initial core rotational energies,
which translate into ratios $E^\textrm{mag}_\ppns/\Erotpns$ orders of
magnitude smaller than in non-CHE models. The large differential
energy reservoir of CHE models is partly tapped into by magnetic
energy, making $E^\textrm{mag}_\ppns/\Erotpns$ comparable to or larger
than for non-CHE models after $t_\textrm{pm}\gtrsim 2\,\sek$
(Fig.\,\ref{Fig:globvars-2} \panel{bottom left}). The magnetic field
strengths on the PNS surfaces, on the other hand, are much stronger
than for the neutrino-driven explosions.  The poloidal component in
particular is consistently around $B^{\mathrm{pol}}_{\mathrm{srf}}
\sim 10^{13} \, \Gauss$.

Based on these results, we suggest that the presence of (moderately)
strong, large-scale magnetic fields in the progenitor star connecting
the PNS surface to and beyond the gain layer is the main reason for
the ability of \modls{35OC}, \modelname{A13}, \modelname{A26},
\modelname{A30} to launch explosions despite their high compactness
and despite not featuring particularly high neutrino heating rates.
It is precisely this subset of progenitors which possesses a
magnetized layer around a mass coordinate of $m\approx 2 \, \msol$
(shaded regions in \figref{Fig:init-angmom}), whereas the absence of
such a layers seems to preclude the possibility of MHD explosions.  We
note that the intermediate case of \modl{A39} exhibits a magnetic
field up to almost this mass coordinate and that the \modl{A17}
accretes a rather narrow shell with non-vanishing magnetic field
around the time when it accelerates a late outflow resembling PM
winds. Thus, we find a connection between the PM activity and the
topology of the magnetic field in the stellar
progenitor. Unfortunately, current stellar evolution models cannot
accurately obtain the topology of the field in the pre-collapse stage
\citepalias[e.g.][]{Aloy__2021__MonthlyNoticesoftheRoyalAstronomicalSociety__MagnetorotationalCoreCollapseofPossibleGRBProgenitorsII.FormationofProtomagnetarsandCollapsars}.
\Modl{A05} produces a similar outflow after the (poloidal) PNS surface
field has grown by several orders of magnitude without accreting a
magnetized layer. This suggests a second way to create the field
configuration necessary to launch such an outflow by transporting
field from the interior of the PNS to its surface or amplifying it
locally.

\subsection{Failed explosion}
\label{sSek:Failexp}

The evolution of \modl{16TI} is affected strongly by its very rapid
rotation.  In the innermost $m \approx 2 \, \msol$, its initial
density profile, the compactness derived from it, and the parameters
$M_4$ and $\mu_4$ roughly fall into the range established by models
exploding due to neutrino heating.  Consequently, the mass accretion
rate is, at least at early times, comparable to those.  At the same
time, centrifugal support reduces the release of gravitational energy
from the matter falling onto the PNS, thus lowering the neutrino
luminosity.  After $\tpb \approx 0.25 \, \sek$, when the entire inner
core has been accreted, this reduction becomes less important, but
then a relatively shallow density profile causes the mass accretion
rate to drop only slowly.  Measured in terms of the mass accretion
rate, the neutrino luminosity never gets high enough to trigger an
explosion with $\zeta \sim 0.4$ for most of the evolution.  Until
$\tpb \sim 2 \, \sek$, neutrino heating is still fairly important in a
gain layer whose mass is shrinking below $M_{\igain} < 10^{-3} \,
\Msol$.  Advection through the gain layer remains faster than neutrino
heating, though only by a factor of about 2.

We note that \modl{A39} launches an explosion under conditions that
are not too dissimilar from the state of failing \modl{16TI}.  The
former model, however, possesses a magnetic field that is sufficiently
strong to aid in the process of reviving the shock wave.  In
\modl{16TI}, on the other hand, once the entire Fe core has fallen
onto the PNS, the ensuing drop of the mass accretion rate goes along
with a transition to the accretion of a non-magnetized shell.  We note
the essentially unmagnetized structure of \modl{16TI} for $1.8\lesssim
m/M_\odot \lesssim 8.8$ compared to the unmagnetized shell surrounding
the core of \modl{A39} for $2\lesssim m/M_\odot \lesssim 4.6$ in
\figref{Fig:init-angmom}.  Hence, the magnetization of the gain layer
drops, eliminating the prospects of an explosion due to a mixed
neutrino-MHD mechanism like in \modl{A39}. Again, the topology of the
magnetic field in the progenitor star is determinant for the fate of
the post-collapse system.

Failing to launch an explosion, $\MPNS$ steadily increases until it
will unavoidably produce a BH.  By $\tpb \gtrsim 3 \, \sek$, the PNS
exceeds the limit for stability for a non-rotating NS.  However, its
very high rotational energy delays the collapse by several seconds and
allows for the accretion of additional mass.  Indeed, \modl{16TI} has
not formed a BH even after $t_\mathrm{pb}=6.1\,\sek$.

\begin{figure}
  \centering
  \includegraphics[width=\linewidth]{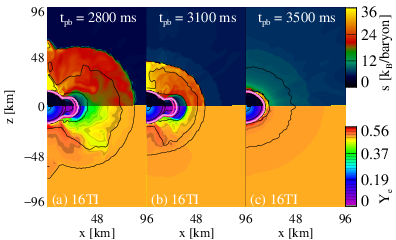}
  \caption{
    Specific entropy and electron fraction (colour maps) together with
    isodensity lines ($\rho = 10^{15,14,13,...} \, \gccm$, black lines)
    and the neutrinospheres (pink lines) of \modl{16TI} around the
    temporary formation of a rotationally supported disc around the PNS.
  }
  \label{Fig:2dPNS16TI}
\end{figure}

In the time interval $\tpb \approx 2.6...3.2 \, \sek$, the layers
located in the progenitor at Lagrangian mass coordinates $m \approx
2.32...2.45 \, \Msol$ are falling onto the PNS.  These layers possess
a particularly high specific angular momentum in excess of our
estimates for the angular momentum corresponding to the last stable
orbit for a maximally rotating Kerr BH encompassing all the matter
enclosed by them (\figref{Fig:init-angmom}; note the yellow dashed
line).  To illustrate the effect of rotation on the structure of the
core, we compare the profiles of $j (m)$ in the equatorial plane for
several times (solid lines in the \panel{top} panel of the figure) to
the specific angular momentum of matter in Keplerian rotation about
the centre, defined as $j_{\mathrm{Kep}} = r^2 \sqrt{-\Phi}$, where
$\Phi$ is the gravitational potential (dashed lines).  Where $j >
j_{\mathrm{Kep}}$, centrifugal forces can stop accretion onto the PNS,
which is the case during this phase.  Consequently, we observe the
formation of a small rotationally supported torus just above the PNS
surface.  As displayed in \figref{Fig:2dPNS16TI} for $\tpb = 2.8$ and
$3.1 \, \sek$, this torus is adjacent to the PNS and extends from its
surface to about twice its radius.  The torus is composed of matter
with densities of $\rho > 10^{11} \, \gccm$ (black contours).  This
high density makes it partially optically thick to neutrinos and the
\nusps (pink lines) encompass large swaths of it.  In the central
regions of the torus, the emission of neutrinos occurs on rapid
timescales compared to the dynamics and dominates over their
absorption.  Consequently, the matter approaches an equilibrium state
characterized by low entropy and low electron fraction, the absorption
equilibrium of
\cite{Just__2021__arXiveprints__NeutrinoAbsorptionandOtherPhysicsDependenciesinNeutrinoCooledBlackHoleAccretionDisks}.

Since the distribution of $j$ shows an abrupt decline toward higher
radii, gas accreted after $\tpb \approx 3.2 \, \sek$ comes with much
lower specific angular momentum.  After only a brief time, the
centrifugal support of the torus no longer suffices to balance gravity
and the ram pressure of the newly accreted gas, and the torus breaks
down and is quickly absorbed by the PNS.  In the subsequent seconds of
the evolution, the specific angular momentum of the accreted gas
remains below the Keplerian value by a factor of a few.  The structure
of the core does not change from the state displayed in the
\panel{right} panel of \figref{Fig:2dPNS16TI}.  Only a very thin layer
of hot gas with high electron fraction separates the PNS from the
infalling matter.  This configuration may only change when the PNS
collapses to a BH or when the next layer with potentially
super-Keplerian rotation, located at a mass coordinate of $m \approx
3.1 \, \Msol$ will be accreted.  Until this point, the PNS can only
sustain itself against its own gravity due to its fast rotation.
Hence, any sufficiently efficient process leading to a loss of angular
momentum to the exterior or a flattening of the differential rotation
(e.g., non-axisymmetric instabilities) might lead to BH collapse
before the accretion of this layer can produce another PNS-plus-torus
system.

We note that we had observed the formation of a qualitatively similar
system in previous models such as \modelname{35OC-Rw},
\modelname{35OC-RRw} or \modelname{35OB-RRw}
\citepalias{Aloy__2021__MonthlyNoticesoftheRoyalAstronomicalSociety__MagnetorotationalCoreCollapseofPossibleGRBProgenitorsII.FormationofProtomagnetarsandCollapsars},
which also show a very extended disc around the PNS.  In that case,
however, the disc formation was an inside-out process driven by the
magnetic transport of angular momentum from the centre of the PNS to
its outer layer, which consequently gains centrifugal support and
expands.  Here, on the other hand, the torus is formed due to the high
angular momentum the gas possesses already in the pre-collapse star.
Hence, contrary to the inside-out mechanism, for which the disc forms
from gas belonging to the PNS, the outside-in process relies on
neutrino emission to drive $Y_e$ to the low values we find in
\modl{16TI}.

\begin{figure}
  \centering
  \includegraphics[width=\linewidth]{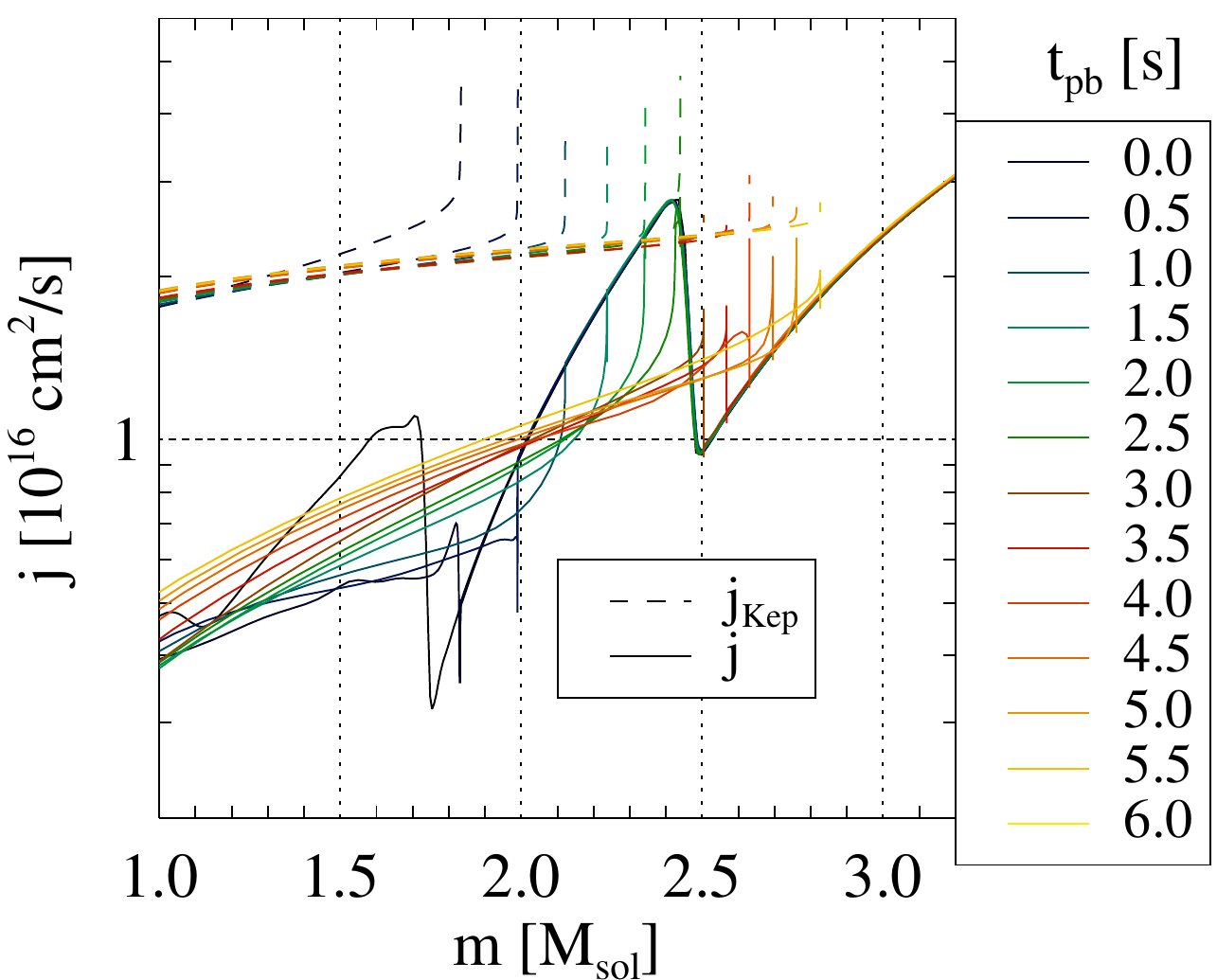}
  \caption{
    Profiles of the actual specific angular momentum in the equatorial plane of
    \modl{16TI}  (solid lines) compared to the
    Keplerian specific angular momentum (dashed lines).
  }
  \label{Fig:2dPNS16TI-j}
\end{figure}

\section{Potential future evolution}
\label{Sek:espekulatius}

\begin{figure}
  \centering
  \includegraphics[width=\linewidth]{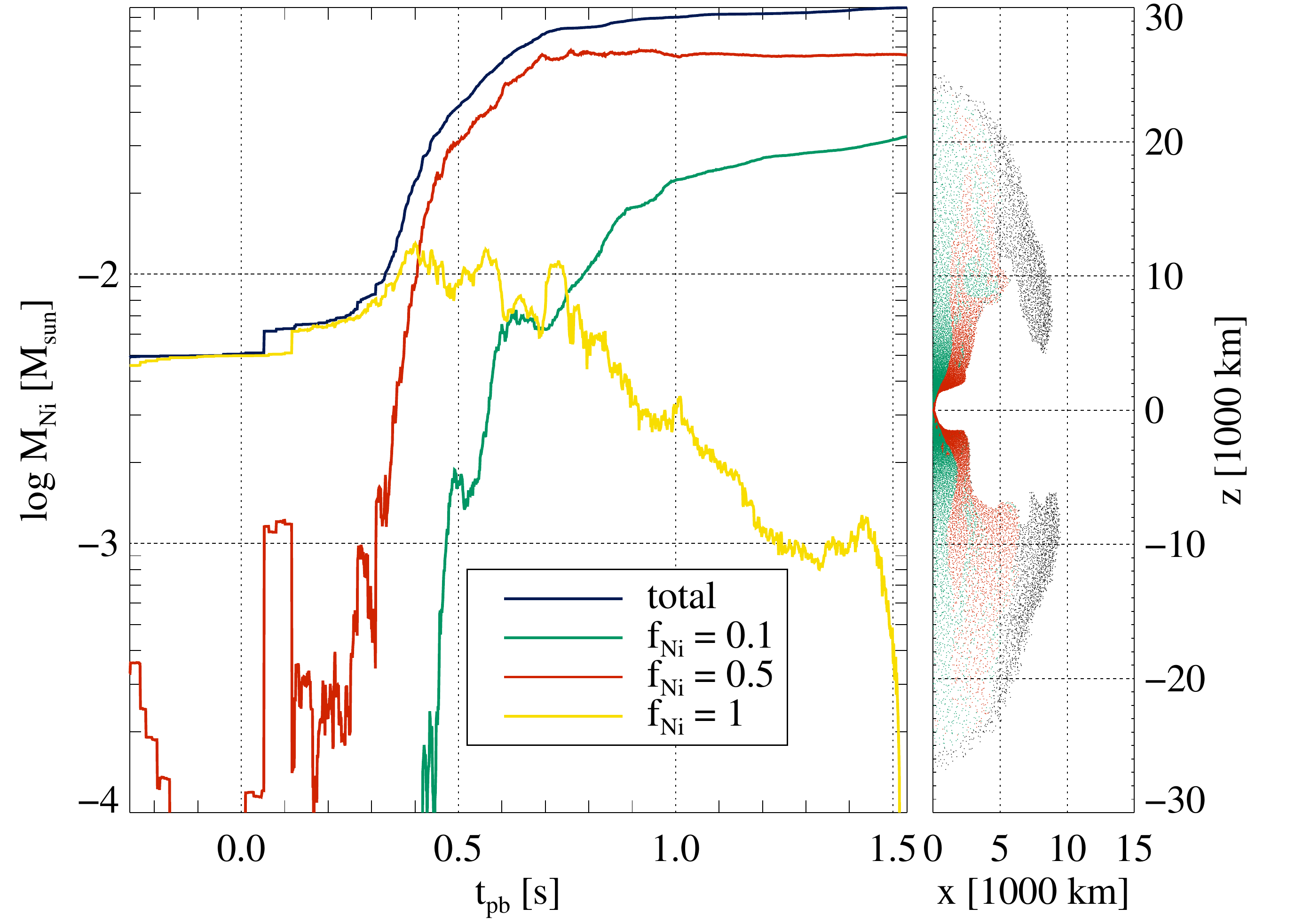}
  \caption{
    \panel{Left panel}: time evolution of the ejected Ni mass of \modl{A30}.
    The dark blue line represents the total Ni mass, and the coloured
    lines show the mass contained in gas with a production factor
    $f_{\mathrm{Ni}}$ as indicated in the legend.
    \panel{Right panel}: spatial distribution of unbound fluid elements at the
    end of the simulation.  Black symbols stand for Lagrangian tracers
    that do not contain any Ni, while tracers with $f_{\mathrm{Ni}} =
    0.1, 0.5, 1$ are displayed by the same colours as in the
    \panel{left panel}.  The yellow Lagrangian tracers are very scarce
    at the end of the computed time and, hence, hardly visible at the
    interface between the jet cavity and the progenitor star.
  }
  \label{Fig:A30Ni}
\end{figure}

Since the dynamics that our models undergo happens deep inside of a
massive star, electromagnetic radiation cannot scape out of the region
of interest and yield an observable signal. Only gravitational waves,
neutrinos, and the nucleosynthetic yields formed at this early stage
of the explosion may inform us about the dynamics of our simulations.
However, large uncertainties beset any attempt to determine the type
of explosion from the state of our models at this point.  In addition
to the diagnostic explosion energies, ejecta geometries and the type
of compact remnant, we aim here at providing very rough estimates of
(i) the mass of radioactive Ni synthesized and ejected, and (ii) of
the early bolometric light curves.

 \subsection{Estimation of the Ni mass}
 \label{sec:Nimass}
 
 A relatively large amount of Ni mass synthesized during the early
 phases of the explosion could explain the high luminosity of SLSNe or
 HNe. As we shall show in this section, this is not the case in our
 models.

 In a manner similar to the flashing scheme that our EOS uses to
 determine the composition of the baryonic component of the gas
 \citet{Rampp_Janka__2002__AA__Vertex,Mueller_et_al__2016__mnras__Asimpleapproachtothesupernovaprogenitor-explosionconnection},
 we assume that a fraction $f_\textrm{Ni} \le 1$ of all lighter
 elements contained in an unbound fluid element of the ejecta is
 immediately burned to Ni once the temperature exceeds a threshold of
 $T_{\textrm{Ni}} = 5 \, \mathrm{GK}$.  At sufficiently high entropy
 per baryon, photodisintegration of Ni into $\alpha$ particles may
 significantly limit the amount of Ni synthesized.  Following
 \cite{Surman_2011ApJ...743..155},
 \cite{Hayakawa_Maeda__2018__apj__ACollapsarModelwithDiskWindImplicationsforSupernovaeAssociatedwithGammaRayBursts},
 we account for this effect by computing the specific entropy of the
 baryons,
\begin{equation}
  \label{Gl:Hayakawa-s}
  s_{\mathrm{b}} =
  \frac{4 a_{\mathrm{rad}} m_{p} T^3}{3 \rho k_{\mathrm{B}}},
\end{equation}
where $a_{\mathrm{rad}}$, $m_{p}$, and $k_{\mathrm{B}}$ are the
radiation constant, the proton mass, and Boltzmann's constant,
respectively.  The Ni fraction depends on $s_{\mathrm{b}}$,
\begin{equation}
  \label{Gl:fNi}
  f_{\mathrm{Ni}} =
  \left\{
    \begin{array}{lll}
      1 & \mathrm{for} & s_{\mathrm{b}} < 1, 
      \\
      0.5 & \mathrm{for} & 1 < s_{\mathrm{b}} < 10,
      \\
      0.1 & \mathrm{for} & 10 < s_{\mathrm{b}}.
    \end{array}
  \right.
\end{equation}
In practice, we record the evolution of Lagrangian tracer particles.
If the temperature of a tracer surpasses $T_{\mathrm{Ni}}$, we assign
it a fraction of Ni according to its entropy.  This fraction can
change as long as the temperature is higher than $T_{\mathrm{Ni}}$.
After cooling below this value, we assume that the composition freezes
out.

Note that our simulations use an EOS that contains, in addition to
baryons, also leptons and photons.  These additional components make
the entropy in our simulations in general greater than the
approximation of
\cite{Hayakawa_Maeda__2018__apj__ACollapsarModelwithDiskWindImplicationsforSupernovaeAssociatedwithGammaRayBursts}.
To use their recipe together with our EOS, we would have to
recalibrate the entropy thresholds to account for the systematic
differences.  Such an improvement of the method would require a
detailed comparison to the conditions for NSE and is beyond the scope
of this article.  We cautionary note that our simple estimates can be
off by several $10 \, \%$.\footnote{For comparison, using a detailed
  nucleosynthesis network,
  \cite{Reichert__2021__MonthlyNoticesoftheRoyalAstronomicalSociety__NucleosynthesisinMagnetoRotationalSupernovae}
  obtained $M_\mathrm{Ni}=0.0473M_\odot$ for \modl{35OC}. For the same
  model, we obtain $M_\mathrm{Ni}=0.048M_\odot$ with the simplified
  method described here. The striking similarity of the results should
  not be taken as a measure of the goodness of the simplified
  treatment. Applied to other models of
  \cite{Reichert__2021__MonthlyNoticesoftheRoyalAstronomicalSociety__NucleosynthesisinMagnetoRotationalSupernovae},
  discrepancies of a few $10\%$ are found.}  Thus, we only use them to
obtain general tendencies within our set of stars.  The last column of
\tabref{Tab:Globerview} ($M_\mathrm{Ni}$) lists the results.

We discuss first a representative case, namely \modl{A30}, for which
\figref{Fig:A30Ni} shows both the time evolution of the Ni mass (left
panel) and the spatial distribution of Ni at the final time (right
panel).  Some of the fluid elements that will be ejected already
contain Ni before bounce, $\tpb < 0$.  At that time, they possess low
entropy and, thus, consist entirely of Ni ($f_{\mathrm{Ni}} = 1$,
yellow line in the \panel{left panel}).  In the first phase after
bounce, most ejecta that surpass $T_{\mathrm{Ni}}$ do so at low
entropy and therefore belong to the same group.  However, as neutrino
heating and shock waves in the jet increase the entropy of the
tracers, some of the Ni is dissociated and $f_{\mathrm{Ni}}$ decreases
for the aforementioned fluid elements. Additional Ni production occurs
mostly in tracers with higher entropy and, thus, low values of
$f_{\mathrm{Ni}} = 0.5$ and $0.1$ (red and green lines, respectively).
At the end of the simulation, gas with $f_{\mathrm{Ni}} = 0.5$
dominates the Ni content of the model, mostly distributed in the
cocoon of the jet (red points in the \panel{right panel}), while the
jet beam contains matter with higher entropy and $f_{\mathrm{Ni}} =
0.1$ (green symbols).  Surrounding the jets we find ejecta that never
reached the threshold temperature for Ni formation.

At first glance, our models synthesize insufficient Ni mass to even
explain typical CCSNe, which yield $\sim 0.1M_\odot$ of Ni. Even less
can they account for more energetic hypernovae, which produce larger
amounts of Ni \citep{Iwamoto_1998Natur.395..672,
  Drout_2011ApJ...741...97, Lyman_2016MNRAS.457..328}. However, we
first note that all our models, save for the non-exploding
\modl{16TI}, show (diagnostic) explosion energies of the order of the
canonical value of $E_{\mathrm{exp}} = 10^{51} \, \erg$ or more (see
\tabref{Tab:Globerview}).  Extrapolating to subsequent phases, further
growth is possible, unless the collapse of the PNS to a BH shuts down
the injection of energy into the surrounding gas.  Should a BH form,
further Ni can be synthesized in the winds emerging from the accretion
disk
\citep[e.g.][]{MacFadyen__2001__apj__Supernovae_Jets_and_Collapsars,
  Kohri_Narayan_Piran__2005__apj__Neutrino-dominated_Accretion_and_Supernovae,
  Kumar_2008MNRAS.388.1729,
  Hayakawa_Maeda__2018__apj__ACollapsarModelwithDiskWindImplicationsforSupernovaeAssociatedwithGammaRayBursts},
that will surround the BH after the formation of a
collapsar. Exploring further this regime is beyond the scope of this
paper. Also, explosive nucleosynthesis due to the interaction of the
accretion disc wind with the stellar envelope may yield some extra Ni
production
\citep{Woosley_Weaver__1995__apjs__The_Evolution_and_Explosion_of_Massive_Stars.II.Explosive_Hydrodynamics_and_Nucleosynthesis,
  Maeda_2009MNRAS.394.1317}. However, this mechanism tends to operate
once the BH is sufficiently massive, as it has swallowed most of the
stellar envelope. That limits the availability of envelope materials
for explosive nucleosynthesis \citep{Maeda_2009MNRAS.394.1317},
specially in CHE models with masses below, say, $8M_\odot$ (i.e.,
\modelname{A05}, and \modelname{A08}). These models may, at most, form
BHs as massive as their initial pre-SN mass minus the ejecta mass.
Neutrino-driven explosions are systematically less violent than
magnetorotational ones.  After a shock revival powered by neutrino
heating, \modls{A05} and \modelname{A17} enter a PM-like phase in
which the explosion energy increases at very high rates.  While all
explosions show an asymmetric geometry, to a certain degree enforced
by the assumption of axisymmetry, a stronger contribution of magnetic
fields to the explosion enhances this tendency.

\subsection{Approximate bolometric light curves}
\label{sec:lightcurves}

In order to gain further insight on the kind of high-energy transient
that our models yield, we have computed approximate analytic light
curves due to the radioactive decay of Ni based upon
\cite{Dado_Dar__2015__apj__AnalyticalExpressionsforLightCurvesofOrdinaryandSuperluminousTypeIaSupernovae}. For
more elaborate models including 3D Monte-Carlo simulations we refer
the reader to, e.g. \cite{Vurm_Metzger_2021ApJ...917...77}. Our
models, furthermore, develop compact cores with large rotational
energy ($\Erotpns >10^{52}\,$erg). Hence, we also include the
contribution of a central energy source that may tap a fraction of the
core's rotational energy, thermalize it with the SN ejecta, and may
modify the early bolometric light curve. We consider separately cases
which do not produce a BH during the computed time (\modelname{A05},
\modelname{A08}, and \modelname{A17}) from the rest of the progenitors
that undergo CHE. Models that do not produce BHs, possess adequate
conditions for hosting a PM, and we refer collectively to them as
proto-magnetar candidates (PMCs). Indeed, one of them has already
shown transient PM spin-down activity (\modl{A17}). Thus, their light
curves may display signatures of a central PM engine on longer terms
than calculated in this work. For the rest of the models, we assume
that the spin-down of the BH constitutes the central engine of a GRB
\citep{Nathanail_2015MNRAS.453L...1, Nathanail_2016MNRAS.455.4479}, in
which part of the BH released energy will be transferred to the SN
ejecta.

\begin{table*}
  \centering
  \begin{tabular}{| l | llllll | llllll |}
    \hline
    Model & $L_{\mathrm{pm}, 47}$ & $B^{\rm pol}_{\mathrm{srf},14}$ & $\bar\Omega$ & $R_{\ppns}$ & $I_{\ppns, 45}$ & $t_{\rm pm}$ & $L_{\mathrm{pk},43}$ & $t_{\rm pk}$  & $t_{\rm r}$ & $\Mej$ & $V_{\mathrm{ej},9}$  & $E_{\rm k, 52}$ \\*[2pt] 
    & [$\ergs$] & [G] & [Hz] & km & [g\,cm$^2$] & [days] & [$\ergs$] & [days] &  [days] & $M_\odot$ & [cm\,s$^{-1}$] & [erg]\\ 
    \hline
    A05   &   2.96 &   3.37 & 2794 &   17.3 &   2.61 &  0.208 &   15.1 &   9.02 &   14.2 &   2.81 &   1.6 &  0.72 \\
    A08   &   35.1 &   7.73 & 3143 &   18.0 &   3.26 & 0.030 &   2.64 &   10.9 &   22.9 &   5.80 &   1.3 &  0.95 \\
    A17   &   28.8 &   4.00 & 3368 &   20.4 &   2.45 & 0.035 &  0.82 &   20.9 &   46.0 &   14.9 &  0.82 &  0.98 \\
    \hline
  \end{tabular}
  \caption{
      Properties of models with potential PM activity: model name,
      engine luminosity, surface average poloidal magnetic field of
      the PNS, representative rotational frequency, PNS equivalent
      spherical radius, PNS moment of inertia, decay time of the
      spin-down luminosity, total luminosity maximum time, total
      luminosity maximum, effective photon diffusion time, ejecta
      mass, ejecta velocity, and ejecta kinetic energy. We use the
      convention $A_x = A/10^x$ to express various quantities in the
      table. A value $f_{\rm er}=0.5$ is used for the calculation of
      $L_{\rm pm}$ and other derived magnitudes. All the properties of
      our models are obtained at $t=t_{\rm f}$.
    }
  \label{Tab:lightcurve}
\end{table*}

\begin{table*}
  \centering
  \begin{tabular}{| l | llllll | llllll |}
    \hline
    Model & $L_{\mathrm{pm}, 47}$ & $B^{\rm pol}_{\mathrm{srf},14}$ & $\bar\Omega$ & $R_{\ppns}$ & $I_{\ppns, 45}$ & $t_{\rm BZ}$ & $L_{\mathrm{pk},43}$ & $t_{\rm pk}$  & $t_{\rm r}$ & $\Mej$ & $V_{\mathrm{ej},9}$ & $E_{\rm k, 52}$\\*[2pt] 
    & [$\ergs$] & [G] & [Hz] & km & [g\,cm$^2$] & [days] & [$\ergs$] & [days] &  [days] & $M_\odot$ & [cm\,s$^{-1}$] & [erg]\\ 
    \hline
    A13   &  0.150 &  0.259 & 2864 &   26.7 &   7.02 &   3.53 &   9.79 &   17.8 &   41.3 &   10.5 &  0.72 &  0.53 \\
    A20   &  0.147 &  0.236 & 5134 &   18.3 &   3.53 &   4.22 &   6.83 &   22.7 &   59.1 &   17.5 &  0.58 &  0.58 \\
    A26   &   1.47 &  0.553 & 4785 &   25.5 &   6.87 &  0.657 &   1.51 &   6.09 &   64.8 &   23.1 &  0.64 &  0.93 \\
    A30   &   1.43 &  0.520 & 4589 &   34.1 &   8.00 &  0.728 &   1.49 &   6.75 &   71.0 &   27.0 &  0.62 &   1.04 \\
    A39   & 0.02 & 0.095 & 3981 &   24.1 &   4.27 &   27.7 &   7.31 &   87.6 &   109 &   36.6 &  0.36 &  0.47 \\
    \hline
  \end{tabular}
  \caption{
    Same as Tab.\,\ref{Tab:lightcurve} but for models with potential
    BH spin-down activity. Instead of listing the decay time of the
    spin-down luminosity, we provide the decay time of the
    Blandford-Znajek process.
  }
  \label{Tab:lightcurvebh}
\end{table*}

The radioactive contribution to the light curve follows exactly the
analytic estimation of
\cite{Dado_Dar__2015__apj__AnalyticalExpressionsforLightCurvesofOrdinaryandSuperluminousTypeIaSupernovae},
taking as inputs from our models the Ni Mass
(Tab.\,\ref{Tab:Globerview}), the ejecta asymptotic velocity, $V_{\rm
  ej}$, and the ejecta mass, $\Mej$. All these quantities are
approximate and not the result of a sufficiently long and detailed
calculation. We estimate an upper bound to the ejecta mass as the
difference between the stellar mass at the brink of collapse, $M_0$,
and the PNS mass at $t_{\rm pb}=t_{\rm f}$
(i.e. $\Mej:=M_0-\MPNS(t_{\rm f})$). We note that $\Mej$ is
significantly larger than the mass unbound at $t_{\rm pb}=t_{\rm f}$
($M_{\rm ej}$; see \figref{Fig:globvars}) or the mass enclosed by the
shock at that time ($M_{\rm sh, e}$; Tab.\,\ref{Tab:Globerview}). At
$t_{\rm pb}=t_{\rm f}$ the explosion is ongoing and both $M_{\rm ej}$
and $M_{\rm sh,e}$ are still growing until they reach their final
values. The asymptotic ejecta velocity is estimated as $V_{\rm
  ej}=\sqrt{2E_{\rm k}/\Mej } = \sqrt{2 (\Erotpns f_{\rm er} + E_{\rm
    exp})/ \Mej}\approx \sqrt{2 \Erotpns f_{\rm er} / \Mej}$, where we
allow for the possibility that only a fraction $f_{\rm er }$ of the
PNS rotational energy can be converted into kinetic energy of the
ejecta, $E_{\rm k}$. The last approximation results from the fact that
the $\Erotpns$ ($>10^{52}\,$erg) is one order of magnitude (or more)
larger than $E_{\rm exp}$ ($<2\times 10^{51}\,$erg; see
Tab.\,\ref{Tab:Globerview}). Typical ejecta velocities in all our
models are $V_{\rm ej}\approx (0.4-1.6)\times
10^{9}\,\text{cm\,s}^{-1}$. The corresponding luminosity reads
\citep{Dado_Dar__2015__apj__AnalyticalExpressionsforLightCurvesofOrdinaryandSuperluminousTypeIaSupernovae}
\begin{align}
L_{\mathrm{nuc}}(t)=\frac{e^{-t^{2} / 2 t_{\mathrm{r}}^{2}}}{t_{\mathrm{r}}^{2}} \int_{0}^{t} x e^{x^{2} / 2 t_{\mathrm{r}}^{2}} \dot{E} \, d x ,
\label{eq:Lnuc}
\end{align}
where 
\begin{align}
t_{\rm r}=\sqrt{ \frac{3\Mej f_{\rm e} \sigma_{\rm T}}{8\pi m_{\rm p} c V_{\rm ej}} },
\label{eq:tr}
\end{align}
$\sigma_{\rm T}$ is the Thompson cross section, $m_{\rm p}$ the proton
mass, $c$ the light speed, $f_{\rm e}\approx 0.275$ is the fraction of
free electrons, and $\dot E=\dot E_\gamma + \dot{E}_{e^+}$, where
\begin{align}
  \dot{E}_{\gamma}=& \frac{M_\mathrm{Ni}}{M_{\odot}}\left[7.78 A_{\gamma}^\mathrm{Ni} e^{-t / 8.76 \mathrm{~d}}\right.\nonumber\\
  &\left.+1.50 A_{\gamma}^\mathrm{Co}\left[e^{-t / 111.27
        \mathrm{~d}}-e^{-t / 8.76 \mathrm{~d}}\right]\right] 10^{43}
  \mathrm{erg} \mathrm{s}^{-1}
\end{align}
is the power supplied by $\gamma$-rays, and
\begin{align}
\dot{E}_{\mathrm{e}^{+}}=& \frac{M_\mathrm{Ni}}{M_{\odot}} A_{\mathrm{e}} \left[e^{-t / 111.27 \mathrm{~d}}-e^{-t / 8.76 \mathrm{~d}}\right] 10^{43} \mathrm{erg} \mathrm{s}^{-1}, 
\end{align}
is the power supply in the form of the kinetic energy of positrons.
$A_{\rm e}\approx 0.05$ is the ratio of the energy released as
positron kinetic energy and as $\gamma$-ray energy in the decay of
$^{56}$Co.  $A_\gamma^\mathrm{Ni}$ and $A_\gamma^\mathrm{Co}$ are the
absorbed fractions of the $\gamma$-ray energy in the SN ejecta from
the decay of $^{56}$Ni and $^{56}$Co, respectively. They are computed
as $A_\gamma \approx 1-e^{-\tau_\gamma}$, with
$\tau_\gamma=3\Mej\sigma_t/(8\pi m_p V^2_{\rm ej}t^2)$, and
$\sigma_t=9.5\times 10^{-26}\,$cm$^2$ ($\sigma_t=8.7\times
10^{-26}\,$cm$^2$) for $^{56}$Ni ($^{56}$Co).

\begin{figure}
  \centering
  \includegraphics[width=\linewidth]{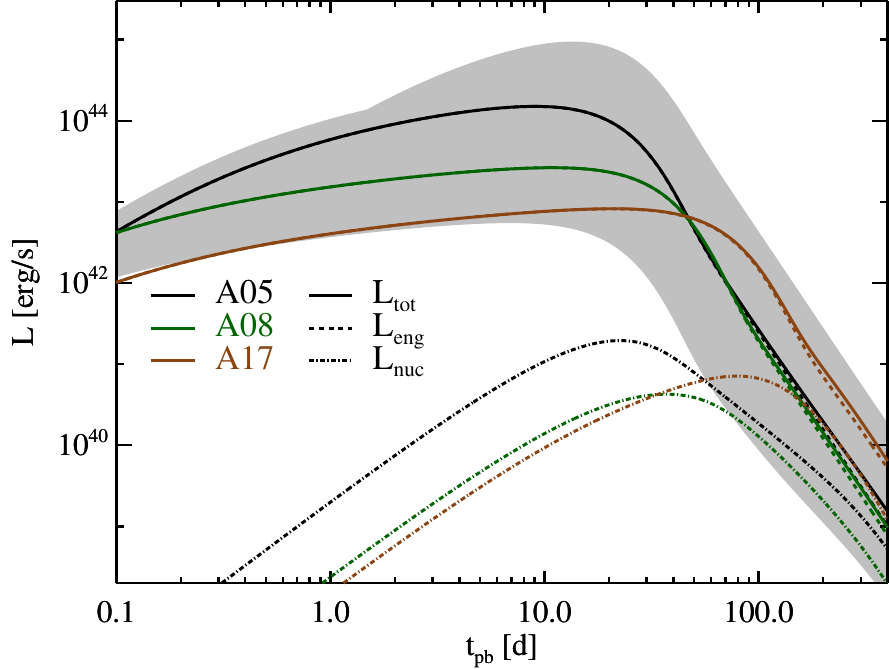}
  \caption{
    Solid lines: Approximate bolometric light curves of models
      that yield successful SNe and have not produced a BH in the
      course of the computed time. Dashed lines: extrapolated light
      curves produced by a PM with the same surface poloidal magnetic
      field and rotational period than the PNS of the models indicated
      in the legends. For both the solid and dashed lines, we assume
      that the ejecta energy is only a fraction $f_{\rm er}=0.5$ of
      the available PNS rotational energy and may eventually produce
      photons that contribute to the light curve. Dash-dotted lines:
      analytic light curves produced by the radio active decay of the
      same mass of Ni as in our models. Grey region: approximate range
      of uncertainty of the predicted bolometric light curve of
      \modl{A05}. The upper (lower) boundary assumes $f_{\rm er}=0.9$
      and $f_{\rm pm}=10$ ($f_{\rm SN}=0.01$ and $f_{\rm pm}=0.1$).
    }
  \label{Fig:lightcurves}
\end{figure}
\begin{figure}
  \centering
  \includegraphics[width=\linewidth]{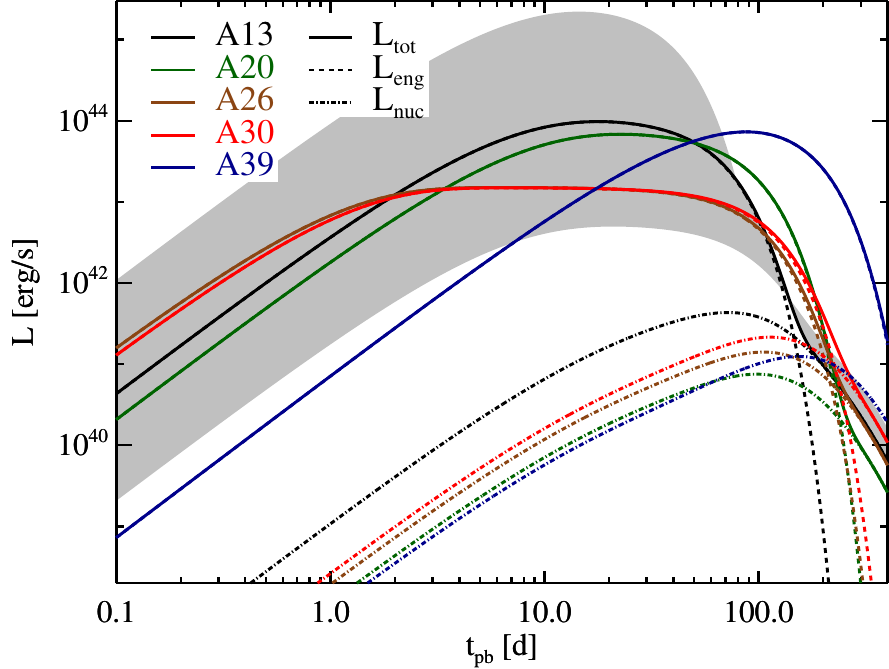}
  \caption{
    Solid lines: Approximate bolometric light curves of models
      that undergo a CHE, yield successful SNe and have produced a BH
      in the course of the computed time. Dashed lines: extrapolated
      light curves produced by the Blandford-Znajek spin-down of a BH
      with the same surface poloidal magnetic field at the outer
      horizon and mass as that of the PNS right before BH
      collapse. For both the solid and dashed lines, we assume that
      the ejecta energy is only a fraction $f_{\rm SN}=0.1$ of the
      available BH rotational energy and may eventually produce
      photons that contribute to the light curve. Dash-dotted lines:
      analytic light curves produced by the radio active decay of the
      same mass of Ni as in our models. Grey region: approximate range
      of uncertainty of the predicted bolometric light curve of
      \modl{A13}. The upper (lower) boundary assumes $f_{\rm SN}=0.9$
      ($f_{\rm SN}=0.01$).
  }
  \label{Fig:lightcurvesbh}
\end{figure}

In order to incorporate the contribution of the PM to the light curves
we follow the same lines as
\citetalias{Kasen_Bildsten__2010__apj__SupernovaLightCurvesPoweredbyYoungMagnetars}. We
assume that the magnetar injects a luminosity due to the spin-down of
the PNS in the ejecta given by
\begin{align}
L_{\rm pm}(t) = \frac{\Erotpns f_{\rm er}}{t_{\rm pm}} \frac{1}{(1+t/t_{\rm pm})^2},
\label{eq:Lspin-down}
\end{align}
where the typical decay time of the spin-down luminosity is
\begin{align}
t_{\rm pm} = \frac{6 I_\ppns c^3}{(B^{\mathrm{pol}}_\mathrm{srf})^2 R_\ppns^6 \bar\Omega^2}.
\label{eq:tspin-down}
\end{align}
To compute $t_{\rm pm}$ we use various inputs taken from the PNS in
each model at $t=t_{\rm f}$ (see Tab.\,\ref{Tab:lightcurve}), namely
the moment of inertia ($I_\ppns$), the equivalent spherical radius
($R_\ppns=(3V_\ppns/4\pi)^{1/3}$; $V_\ppns$ is the PNS volume), the
surface averaged poloidal magnetic field, and a representative
rotational frequency ($\bar\Omega=\sqrt{2\Erotpns/I_\ppns}$). As a
result, we obtain values of $t_{\rm pm}$ significantly smaller than
the effective photon diffusion time (compare $t_{\rm pm}$ and $t_{\rm
  r}$ in Tab.\,\ref{Tab:lightcurve}). These small values of $t_{\rm
  pm}$ contrast with the ones of $\sim 1\,$yr employed in
\citetalias{Kasen_Bildsten__2010__apj__SupernovaLightCurvesPoweredbyYoungMagnetars}
or those of \cite{Nicholl_2017ApJ...835L...8} (who estimated $t_{\rm
  pm}\sim 5\,$days for the Gaia16apd SLSN).  The smaller values of
$t_{\rm pm}$ in our case stem from the typically larger surface
average magnetic fields, PNS radii and PNS inertia moments, and the
fact that the typical periods of the PNSs at hand are $\sim 2\,$ms.

Finally, we compute the contribution to the light curve of the PM central engine as
\begin{align}
L_{\mathrm{eng}}(t)=\frac{e^{-t^{2} / 2 t_{\mathrm{r}}^{2}}}{t_{\mathrm{r}}^{2}} \int_{0}^{t} x e^{x^{2} / 2 t_{\mathrm{r}}^{2}} L_{\rm pm}(x) \,d x .
\label{eq:Leng}
\end{align}

In expression \eqref{eq:Leng} we are implicitly assuming that the
engine's luminosity is efficiently reprocessed into thermal radiation,
and that this is injected at the inner edge of the ejecta. This is a
common assumption in the literature
(\citetalias{Kasen_Bildsten__2010__apj__SupernovaLightCurvesPoweredbyYoungMagnetars};
\cite{Dessart_et_al__2012__mnras__Superluminoussupernovae:$56$Nipowerversusmagnetarradiation,
  Inserra_2013ApJ...770..128,
  Metzger_et_al__2015__mnras__Thediversityoftransientsfrommagnetarbirthincorecollapsesupernovae,
  Nicholl_2017ApJ...835L...8}). The total bolometric light curve is
then
\begin{align}
L_{\rm tot}(t) = L_{\rm eng}(t) + L_{\rm nuc}(t).
\label{eq:Ltot}
\end{align}
We display in Fig.\,\ref{Fig:lightcurves} the bolometric light curves
of PMCs along with the separate contributions of $L_{\rm eng}(t)$ and
$L_{\rm nuc}(t)$.  The contribution of the central PM engine dominates
the light curve until well beyond the luminosity peak, $L_{\rm
  pk}$. The smaller the progenitor mass, the sooner the luminosity
reaches its maximum value and the larger is $L_{\rm pk}$. To
understand this behaviour, we first notice that there is a direct
correlation between the progenitor mass and the ejecta mass. Since all
models develop compact cores with masses clustering in the range
$M_{\rm r}\simeq 2.2M_\odot - 3M_\odot$, and $\Mej:=M_0-\MPNS(t_{\rm
  f})$, the ejecta mass is roughly proportional to the stellar
progenitor mass in our simple light curve model. The larger peak
luminosity for less massive progenitors comes from the fact that
$L_{\rm pk}\sim \Erotpns t_{\rm pm} f_{\rm er} / t_{\rm r}^2 \propto
\Mej^{-3/2}$
\citepalias{Kasen_Bildsten__2010__apj__SupernovaLightCurvesPoweredbyYoungMagnetars}. The
earlier peak time associated to the smaller progenitor mass stems from
the dependence of the effective diffusion time with the ejecta mass,
$t_{\rm r}\propto \Mej^{1/2}$ (Eq.\,\ref{eq:tr}). At this point it is
convenient to reiterate that the ejecta mass that we are using is
significantly larger than $M_{\rm ej}$ and $M_{\rm sh,e}$ (especially
for \modl{A17}, though $M_{\rm sh,e}\approx \Mej$ for
\modelname{A05}). Should the true ejecta mass be smaller than our
upper bound, our models would produce even brighter light curve peaks
at earlier times.
The highest possible luminosities in an engine driven light curve
happen when $t_{\rm pm}\sim t_{\rm r}$
\citepalias{Kasen_Bildsten__2010__apj__SupernovaLightCurvesPoweredbyYoungMagnetars}. This
is not the case in our PMCs, where typically $t_{\rm r}\gg t_{\rm
  pm}$. Hence, we expect that the peak time be $t_{\rm pk}\sim t_{\rm
  pm}/(\ln{(t_{\rm r}/t_{\rm pm})}-1)^{1/2}\sim 0.4t_{\rm pm}$
\citepalias{Kasen_Bildsten__2010__apj__SupernovaLightCurvesPoweredbyYoungMagnetars}. The
relation approximately applies to PMCs, as can be checked from
Tab.\,\ref{Tab:lightcurve}.

Given the strong dependence of the light curve on the spin-down
timescale and on the fraction of the engine's energy that may be
transferred to the ejecta, we roughly quantify the uncertainty in our
approximate light curves as follows. First, we allow for the
possibility that $t_{\rm pm}$ be a factor $f_{\rm pm}$ longer or
shorter than the estimation of \eqref{eq:Lspin-down}. This is
justified because the conditions of the PNS at $t_{\rm pb}=t_{\rm f}$
may experience some evolution on longer time scales
\citepalias{Aloy__2021__MonthlyNoticesoftheRoyalAstronomicalSociety__MagnetorotationalCoreCollapseofPossibleGRBProgenitorsII.FormationofProtomagnetarsandCollapsars},
and also be affected by the fallback of matter not unbound during the
explosion. Second, we anticipate that not all the rotational energy
may be transferred to the ejecta.  Also, as the rotational energy of
the PNS decreases below $\sim \zehn{50}\,\ergs$ the energy loss may
decrease
\citep{Bucciantini_et_al__2009__mnras__Magnetizedrelativisticjetsandlong-durationGRBsfrommagnetarspin-downduringcore-collapsesupernovae}. The
PM injected energy can produce a collimated relativistic jet if the
magnetic to total energy ratio is large enough
\citep{Bucciantini_et_al__2007__mnras__Magnetar-driven_bubbles_and_the_origin_of_collimated_outflows_in_GRBs}. In
this case, the jet may penetrate the SN ejecta without releasing most
of its energy in them, instead producing a GRB
\citep{Bucciantini_et_al__2008__mnras__Relativistic_jets_and_lGRBs_from_the_birth_of_magnetars}.
The upper boundary of the grey-shaded area in \figref{Fig:lightcurves}
corresponds to \modl{A05} using an spin-down time scale $f_{\rm
  pm}=10$ longer than the estimation of \eqref{eq:Lspin-down} ($t_{\rm
  pm}\approx 0.2\,$days) and assuming that most of the PNS rotational
energy is efficiently transferred to the SN ejecta ($f_{\rm
  er}=0.9$). The lower boundary of the uncertainty band results from
taking a 10 times shorter spin-down time scale ($f_{\rm pm}=0.1$) and
a relatively low efficiency for the energy transfer ($f_{\rm
  er}=0.1$).
Even under these assumptions, the light curve of \modl{A05} has a
relatively large (engine-driven) peak luminosity $\sim
\text{few}\times 10^{42}\,\ergs$.

There is a factor $\sim 3$ between the largest and the smallest Ni
mass of PMCs (Tab.\,\ref{Tab:Globerview}). Due to the proportionality
between $L_{\rm nuc}$ and $M_{\rm Ni}$, this radioactive mass
difference reflects in the peaks of $L_{\rm nuc}$ in our
models. Besides this factor, the contribution of the radioactive Ni
decay is only observable well beyond $t_{\rm pk}$ if the spin-down
timescale is significantly shorter than the estimation of
\eqref{eq:tspin-down} (see, e.g., how the black dash-dotted line
penetrates the grey-shaded region after $\sim 50\,$days). We
anticipate that other radioactive isotopes (e.g., $^{66}$Cu), if they
are abundant enough, could increase the radioactive luminosity peak
and the exact post-peak decay slope of the light curve. We defer the
analysis of this possibility to a more detailed study of the nuclear
yields of our models.  Nevertheless, for the typical parameters
employed to compute the light curve here, the decay of the central
engine contribution, rather than the radioactive activity, largely
sets the evolution post-peak.

To incorporate the central energy contribution in BH forming
  models, we assume that the BHs spins down due to the action of the
  Blandford-Znajek (BZ) process.  A fraction of the spin-down energy
  boosts a relativistic jet, which penetrates the ejecta and yields a
  GRB. Another fraction of the energy released, $f_{\rm SN}$, hits the
  ejecta and heats it up (in a process analogous to the jet/ejecta
  interaction shown in
  \citet*{Cuesta_et_al__2015__mnras__Numericalmodelsofblackbody-dominatedgamma-raybursts-IHydrodynamicsandtheoriginofthethermalemission};
  \citet{Cuesta_et_al__2015__mnras__Numericalmodelsofblackbody-dominatedgamma-raybursts-IIEmissionproperties}. We
  note that the parameter $f_{\rm SN}$ is equivalent to $f_{\rm er}$
  in the modeling of the light curves of PMCs. Adapting the results of
  \cite{Nathanail_2015MNRAS.453L...1}, the energy released in the SN
  ejecta due to the spin-down of a BH of mass, $M_{\rm bh}$, with a
  magnetic field, $B_{\rm bh}$, threading its ergosphere is
\begin{align}
L_{\rm BZ}(t) = \dot{E}_{0} e^{-t/t_{\rm BZ}},
\label{eq:LBZ}
\end{align}
where, the spin-down e-folding time is 
\begin{align}
  t_{\rm BZ} &\equiv \frac{3c}{16 G^2 B_{\rm bh}^2 M_{\rm bh} }
  \nonumber \\ &\approx 1.2\,\text{days}\,\left(\frac{B_{\rm
        bh}}{\zehnh{5}{13}\,\text{G}}\right)^{-2}\left(\frac{M_{\rm
        bh}}{2\,M_\odot}\right)^{-1} .
\label{eq:tBZ}
\end{align}
We estimate $\dot E_0$ from the rotational energy of the compact core
at $t_{\rm pb}=t_{\rm f}$, assuming that a proxy to the BH mass is the
mass of the compact remnant at the same time (i.e., $M_{\rm bh}=M_{\rm
  r}(t_{\rm f})$), and that the typical magnetic field strength in the
BH ergosphere is $B_{\rm bh}\sim B_{\rm srf}^{\rm pol}$.  Hence, $\dot
E_0 = f_{\rm SN}\Erotpns / t_{\rm BZ}$. As in the PM case, we compute
the contribution to the light curve of the BH central engine using
\begin{align}
L_{\mathrm{bh}}(t)=\frac{e^{-t^{2} / 2 t_{\mathrm{r}}^{2}}}{t_{\mathrm{r}}^{2}} \int_{0}^{t} x e^{x^{2} / 2 t_{\mathrm{r}}^{2}} L_{\rm BZ}(x) \,d x .
\label{eq:LengBH}
\end{align}
As a reference value, we take $f_{\rm SN}=0.1$ (dashed lines in
\figref{Fig:lightcurvesbh}), but we have considered two other values
to provide a rough range of uncertainty in the predicted light
curve. Precisely, the grey-shaded area in \figref{Fig:lightcurvesbh},
computed for \modl{A13}, is bounded by the cases in which most of the
BH spin-down power is transferred to the ejecta ($f_{\rm SN}=0.9$;
upper boundary of the shaded area) and, by the case in which most of
the energy fuels a GRB jet ($f_{\rm SN}=0.01$; lower boundary). These
two cases may be correlated with the opening angle of the GRB-jet. The
smaller the jet opening angle, the most likely it can penetrate the
ejecta without heating it up appreciably
\citep[cf.][]{Bromberg_et_al__2011__apj__ThePropagationofRelativisticJetsinExternalMedia,
  Mizuta_Ioka__2013__apj__OpeningAnglesofCollapsarJets,
  Aloy__2018__MonthlyNoticesoftheRoyalAstronomicalSociety__OntheExistenceofaLuminosityThresholdofGRBJetsinMassiveStars}. The
variation of the free parameter $f_{\rm SN}$ also indirectly accounts
for the fact that the time dependence of the BZ process included
through \eqref{eq:LBZ} is only strictly valid for BHs with moderate or
small dimensionless spin \citep[for $a\approx 1$, large deviations in
the BZ power can be found; see,
e.g.,][]{Tchekhovskoy_2010ApJ...711...50,
  Mahlmann_2018MNRAS.477.3927}). More elaborate models, which are not
restricted to a range of BH spins \citep{Okamoto_1992MNRAS.254..192,
  Lee_2000PhR...325...83},\footnote{These models neglect the
    possible evolution of the magnetic flux at the BH horizon induced,
    e.g., by the accretion of magnetized mass. Indeed, looking at the
    alternated pattern of magnetized and non-magnetized stellar shells
    that may be accreted onto the BH, the BZ mechanism may be
    intermittent and yield variability in the Poynting outflow
    \citep{Tchekhovskoy_2015MNRAS.447..327}. But even under these
    conditions, the BZ process may operate with a relative high
    efficiency \citep{Mahlmann_2020MNRAS.494.4203}.}  also show a
roughly exponential decay of the BZ luminosity.  In any case, a
temporal decay significantly faster than $\sim (1+t/t_{\rm pm})^{-2}$
assumed for the PM spin-down. The faster decay of $L_{\rm BZ}(t)$
compared to $L_{\rm pm}(t)$ reflects in the faster decay of the light
curve past its peak value in potentially collapsar forming models (to
which we may refer as proto-collapsars; PCs) compared to PMCs (compare
Figs.\,\ref{Fig:lightcurves} and \ref{Fig:lightcurvesbh}). Indeed, the
decay of the light curve past the peak is so fast that the
contribution of the radioactive Ni decay dominates the total
luminosity after a time $t_{\rm pb}\sim 10 \, t_{\rm pk}$ (more in the
case of \modls{A26} and \modelname{A30}). We critically note that a
faster light-curve decline may happen if the breaking index, $m=\ddot
{\bar\Omega} \bar \Omega/\dot{\bar\Omega}^2$ of the PM is
significantly smaller than $m=3$ (which corresponds to the spin-down
of a vacuum magnetic dipole). That would make more difficult
distinguishing observationally among superluminous events produced by
PMCs and PCs on the basis of the decline of the light-curve.

\subsection{Extrapolated high energy transient
 \label{sec:extrapol_transient}}

Considering together the Ni masses, the diagnostic explosion energies
listed in \tabref{Tab:Globerview} and the extrapolated bolometric
light curves of the previous section, we may cautiously suggest the
type of high-energy transient that our models may yield. All PMCs
display $L_{\rm pk}\sim \zehn{43\ldots 44}\,\ergs$, i.e., their
luminosity is broadly compatible with that of SLSNe. Admittedly, there
are (significant) uncertainties in these predictions. The PM breaking
index, value of $t_{\rm pm}$, and the efficiency of conversion of
rotational energy into ejecta luminosity are simple
estimates. However, we have chosen conservative values of the free
model parameters $f_{\rm er}$ and $f_{\rm pm}$ and approximately
quantified the uncertainties of the light curve and, at minimum, all
PCs would produce rather energetic SNe.  For instance, \modl{A17}
produces an incipient PM already at $t_\mathrm{pb}\sim 3\,\sek$ and
displays an extremely fast-growing diagnostic explosion energy
(\figref{Fig:globvars}). Thus, this model may produce an energetic SN,
although in the lower luminosity range of SLSNe ($L_{\rm
  pk}^{\modelname{A17}} \approx \zehn{43}\,\ergs$), with spectral
properties compatible with Type Ic SNe, since the amount of Helium in
the outer layers of the star is very small. Noteworthy, the kinetic
energy of the ejecta in \modl{A17} ($E_{\rm
  k}^{\modelname{A17}}\approx \zehn{52}\,\erg$) is larger than in any
other PMC. The combination of high ejecta kinetic energy and $\sim 1$
order of magnitude smaller peak luminosity than usual SLSNe is typical
of hypernovae, extreme SNe often observed along with GRBs
\citep{Moriya_et_al__2018__ssr__SuperluminousSupernovae}. Should an
ultrarelativistic outflow form (and produce an associated GRB), it
might be launched from the PM some time after the computed evolution
in this paper, when the close vicinity of the PNS reduces its density.
\Modl{A05} might evolve along similar lines, but with a larger
luminosity than \modl{A17}, likely qualifying it as a
SLSN. \Modl{A08}, which also may leave behind an NS rather than a BH,
would be characterized by an SLSN ($L_{\rm pk}^{\modelname{A08}}
\approx \zehnh{2.6}{43}\,\ergs$). Again, as in the case of \modl{A17},
\modl{A08} gathers suitable conditions for the launching of GRB-jet
driven by the central PM. Its smaller PNS rotational energy
($\Erotpns^{\modelname{A08}}\approx 2\times 10^{52}\,\erg$)
anticipates a less energetic relativistic transient than in
\modl{A17}. Thus, for PMCs, the expected observational signature is in
line with the predictions of \AguiDet.

\Modls{A13} and \modelname{A20}, for which \AguiDet predicts a SLSN,
along with a PM (or without a central engine in the case of
\modelname{A13}), may indeed produce such a bright SN by the action of
its central engine, albeit the later is a BH, instead of a NS.  For
higher stellar masses, \modls{A26} and \modelname{A30} display very
early peaks ($t_{\rm pk}<7\,$days), where the luminosity is marginally
compatible with that of a SLSN, followed by a plateau that extends for
a few months. These models have the compact cores with the largest
rotational energy among the CHE models, and among the largest
diagnostic explosion energies. These two facts ally to yield the
largest ejecta kinetic energies of all PCs ($E_{\rm k}\sim
\zehn{52}\,\erg$). Having peak energies $\sim 1$ order of magnitude
smaller than the rest of the PCs and high kinetic energies, these
models may be hypernova candidates (more than SLSNe). Finally,
\modl{A39}, a borderline case between magneto-rotational and
neutrino-driven explosion, features a large peak luminosity ($L_{\rm
  pk}^{\modelname{A39}}\approx \zehnh{3}{43}\,\ergs$) in spite of its
low central engine initial luminosity ($L_{\rm
  pm}^{\modelname{A39}}\approx \zehnh{2}{45}\,\ergs$). The smaller
surface average poloidal magnetic field of all the models in
Tabs\,\ref{Tab:lightcurve} and \ref{Tab:lightcurvebh} is the reason
for its long decay time, $t_{\rm BZ}^\modelname{A39}\approx 28\,$days,
indeed the longest of all CHE models. This decay time is only $\sim 4$
times smaller than its own effective diffusion time scale $t_{\rm r}$,
compared to, e.g., $t_{\rm r}/t_{\rm BZ} \sim 100$ in \modls{A26} and
\modelname{A30}, which favours a rather bright transient
\citep{Arnett__1982__apj__TypeIsupernovaeI-Analyticsolutionsfortheearlypartofthelightcurve}.

Among the BH-forming stars, \modl{16TI} represents the clearest case
for a collapsar.  While it fails to produce an SN (at least within
$t_{\rm pm}<6.1\,\sek$), the other stars from this group may produce
rather bright explosions, if the effect of the central engine is
properly accounted for in our extrapolated light curves
(Sect.\,\ref{sec:lightcurves}).  Given their progenitor structure,
which enables the possibility of forming an accretion disc, BH forming
models could develop a collapsar within a few seconds after $t_{\rm
  f}$.

\section{Conclusions}
\label{Sek:Concl}

Rotation and magnetic fields are common ingredients to explain some of
the most extreme classes of stellar core collapse, \viz SLSNe and
GRBs.  Single massive stars serving as progenitors of CCSNe may only
develop (fast) rotation and (strong enough) magnetic fields under
special evolutionary scenarios.  \AguiDet explored such a possibility
in stellar-evolution models of stars with enhanced rotational mixing
leading to CHE, which ended their hydrostatic burning phases as
Wolf-Rayet stars.  Based on approximate methods, they found that
several of their models with masses between $M_{\textsc{ZAMS}} = 5 \,
\msol$ and $M_{\textsc{ZAMS}} = 39 \, \msol$ might explode as SLSNe or
as GRBs powered by PMs or collapsars.

We used self-consistent two-dimensional numerical simulations of eight
of their progenitors to check these estimates and constrain likely
endpoints of the evolution.  To this set, we added two rapidly
rotating progenitors from
\cite{Woosley_Heger__2006__apj__TheProgenitorStarsofGamma-RayBursts},
\modl{35OC} with $M_{\textsc{ZAMS}} = 35 \, \Msol$ from our previous
studies and \modl{16TI} ($M_{\textsc{ZAMS}} = 35 \, \Msol$) that has
been employed in various studies of the propagation of jets produced
by GRB engines through stellar envelopes.  The simulations included
special relativistic MHD, an approximately general relativistic
gravitational potential, spectral two-moment neutrino transport, and
the relevant reactions between neutrinos and matter.

We are able to run the simulations for very long times (in some cases
to $t_{\rm pb}\gtrsim 6\,s$). However, even longer run times would be
desired, e.g., to go beyond the formation of a BH, which is currently
not feasible with the same approach to the detailed modelling of the
microphysics.  Furthermore, the assumption of axisymmetry might
artificially restrict the dynamics of the stellar cores, in particular
when it comes to possible spiral modes of the PNS and the post-shock
region
\citep{Ott_et_al__2005__apjl__One-armed_Low-TW_Spiral_Instability,Blondin_Shaw__2007__apj__Linear_Growth_of_Spiral_SASI_Modes_in_CCSNe,Fernandez__2010__apj__TheSpiralModesoftheStandingAccretionShockInstability},
the amplification of the magnetic field and dynamos
\citep{Endeve_et_al__2012__apj__TurbulentMagneticFieldAmplificationfromSpiralSASIModes:ImplicationsforCore-collapseSupernovaeandProto-neutronStarMagnetization,Moesta_et_al__2015__nat__Alarge-scaledynamoandmagnetoturbulenceinrapidlyrotatingcore-collapsesupernovae},
and the stability of magnetically driven outflows
(\cite{Mosta_et_al__2014__apjl__MagnetorotationalCore-collapseSupernovaeinThreeDimensions};
\citetalias{Obergaulinger__2021__mnras__MagnetorotationalCoreCollapseofPossibleGRBProgenitorsIII.ThreeDimensionalModels}).
Multi-dimensional simulations may also hold the key to removing some
of the biggest uncertainties regarding the progenitor, \viz the
structure of the magnetic field, in particular in convective layers.

All of our models, save for \modl{16TI}, eventually produce
explosions.  We find shock revival due to neutrino heating and
explosions driven by the magnetorotational mechanism.  The latter
class of explosion is possible if the PNS and the gain layer possess a
strong magnetic field.  Whether or not this condition can be fulfilled
depends on the distribution of the magnetic field in the progenitor
star (in agreement with the findings of
\cite{Bugli__2020__MonthlyNoticesoftheRoyalAstronomicalSociety__TheImpactofNonDipolarMagneticFieldsinCoreCollapseSupernovae};
\citetalias{Aloy__2021__MonthlyNoticesoftheRoyalAstronomicalSociety__MagnetorotationalCoreCollapseofPossibleGRBProgenitorsII.FormationofProtomagnetarsandCollapsars}).
The stellar evolution calculations account for magnetic fields only in
radiative layers and neglect them in convective zones.  Hence, the
gain layer may possess a strong field when a radiative zone is falling
through the shock wave, while being only weakly magnetized during the
accretion of a convective layer.  This connection with the stellar
profile makes MHD explosions viable for the stars from \AguiDet with
ZAMS masses of $M_{\textsc{ZAMS}} = 13, 26,$ and $30 \, \msol$,
besides our \modl{35OC}. In the other models, shock revival is
primarily or entirely driven by neutrino heating and non-spherical gas
flows.

The large masses of the stellar cores translate in very late
explosions in some of the models, in particular at the lower end of
the mass range.  Once launched, the explosions assume a polar geometry
with a more or less pronounced north-south asymmetry.  The diagnostic
explosion energies are moderate to high with maximum values of up to
$E_{\mathrm{exp}} \approx \zehnh{1.9}{51} \erg$, though
$E_{\mathrm{exp}}$ is still growing by the end of the simulations in
most models.  If only a fraction of the PNS or BH rotational energy
can be tapped and transferred to the SN ejecta on longer timescales
than computed here, the implied explosion energies can reach typical
hypernova values.

Before and after the explosion, mass accretion increases the masses of
the PNS, in some cases driving it above the limit for gravitational
stability.  Rotation can allow for the PNS to reach a mass close to
$\MPNS \approx 3 \, \msol$ before collapse.  Our simulations end at BH
formation or after a few seconds of post-bounce evolution.  Hence, our
prospects to confirm or disprove the predictions of \AguiDet regarding
the fate of the stars are limited. However, the expected neutron star
masses in models that do not form BHs exceed $2M_\odot$, categorising
them as massive neutron stars according to
\cite{Antoniadis_2016arXiv160501665A}.

Although the number of models that we have explored here is very small
compared to the large sample of netrino-driven explosions conducted in
\cite{Ertl_et_al__2016__apj__ATwo-parameterCriterionforClassifyingtheExplodabilityofMassiveStarsbytheNeutrino-drivenMechanism},
we find that the two-parameter criterion for the remnant prediction
only marginally applies for rotating, magnetized and low-metallicity
stars. BH forming models typically stem from progenitor stars with
$\mu_4>0.12$, but \modl{A20} challenges this rough division. Instead,
a one parameter division according to $\xi_{2.5}$ appears as a simple
criterion to distinguish between BH forming ($\xi_{2.5}\ge 0.21$) and
PNS forming cases ($\xi_{2.5}< 0.2$).  Likewise, a division based upon
the rotational energy contained within $M_4$ allows for a rough
classification of the potential remnants: low rotational energies in
the pre-collapse stellar matter with $s\le 4$ ($\le
\zehnh{1.4}{48}\,$erg) produce PNSs as compact remnants.  We have
found a dimensionless parameter, which informs about the explosion
type as a function of the magnetic structure in the stellar
progenitor, the ratio of the mean poloidal magnetic field length
within the inner $2.5\,M_\odot$ to the radius of the mass shell
$m=2.5\,M_\odot$, $\lambda_{2.5}$. Neutrino-driven explosions result
if $\lambda_{2.5}<0.1$, while magneto-rotational or mixed type
explosions occur otherwise.
However, we note that the results of this paper apply to axisymmetric
models. In 3D the accretion rate onto the PNS may be significantly
lowered and the prospects to produce a NSs (on longer terms) increase
\citepalias{Obergaulinger__2021__mnras__MagnetorotationalCoreCollapseofPossibleGRBProgenitorsIII.ThreeDimensionalModels}. More
systematic studies, including finer grids of models and extended to
3D, will be the subject of future work.

The clearest indication for PM activity can be found in models
\modelname{A17} and \modelname{A08} where a rapidly rotating and
strongly magnetized PNS with too little mass for a BH to form
accelerates an MHD wind and injects energy at a high rate into the
ejecta.  If the model maintained its structure for a substantial time,
it would, as suggested by \AguiDet, power a PM-driven GRB.  The
explosion energies of the two low-mass models, \modelname{A05} and
\modelname{A08}, do not reach particularly high values by the end of
the simulation. However, given the rates at which they increase and
the strong surface fields of the PNS hint at the possibility of
additional magnetorotational energy input on longer timescales.
Whether this prospect manifests in the form of an SLSN or a GRB cannot
be definitely answered without much longer simulation times.

In all other cases, the PNS collapses to a BH within a few seconds.
At their formation, the BHs rotate at only moderate rates with initial
spin parameters between $a \approx 0.3$ and $a \gtrsim 0.5$.  Mass
accretion continues at rates of several $0.01 \, \msol \isek$.  Thus,
within several hundreds of seconds, the shells at mass coordinates at
$m \approx 3 \msol$ will fall towards the centre.  These shells
possess a specific angular momentum around $j \gtrsim \zehn{16}
\cmcmis$, which would allow for the collapsar-driven GRBs surmised by
\AguiDet.  The point at which, according to our simulation, the disk
would form, is in good agreement with the estimates shown in
\figref{Fig:init-angmom}, in particular the one marked by the blue
asterisk.

The potential for a collapsar-driven GRB is particularly high for
\modl{16TI}.  Accreting rapidly rotating gas which adds centrifugal
support, the PNS grows past the maximum mass for a non-rotating, cold
PNS.  During an intermittent phase, a shell with super-Keplerian
rotation falling onto the PNS forms a geometrically and optically
thick torus around the PNS.  Inside of the torus, neutrino emission
reduces the electron fraction and the entropy.  After a few 100\,ms,
the accretion of gas with lower angular momentum crushes the torus.
The rotational profile of the star makes it likely that at a later
time, around or briefly after the formation of a BH, a torus of
similar characteristics will and allow for collapsar activity.

We have extrapolated our results for models that form PNSs, which we
call PMCs following the naming convention in
\citetalias{Aloy__2021__MonthlyNoticesoftheRoyalAstronomicalSociety__MagnetorotationalCoreCollapseofPossibleGRBProgenitorsII.FormationofProtomagnetarsandCollapsars},
assuming that the rotational energy in the compact remnant may be
tapped by the PM and contribute to the light curve of the SN explosion
at later times. The radioactive decay of Ni produced in our models
during the computed post-bounce time also contributes to the light
curve. Employing an analytic, one-zone model for the bolometric light
curve, we find that our PMCs yield peak luminosities in the range of
SLSNe or, at least, bright SNe ($L_{\rm pk}\gtrsim
\zehn{43}\,\ergs$). Likewise, we have incorporated the contribution of
a BH as a central engine which releases part of its energy into the SN
ejecta in PC models. Very luminous SNe result even if the ejecta only
receives a small fraction ($f_{\rm SN}\sim 1\%$) of the rotational
energy of the BH. Peak luminosities in the range of SLSNe (say,
$L_{\rm pk}\gtrsim \zehn{43\ldots 44}\,\ergs$) are in reach of our
models for more efficient energy transfer to the ejecta ($f_{\rm
  SN}\sim 10\%$). The central engine contribution dominates the
bolometric light curves in all the models analyzed, at least during
the first $\sim 100\,$days of the ejecta evolution. BH forming models
have a light curve with a steeper (exponential) decay after its peak
value, which helps to uncover the radioactive contribution $\sim
200\,$days after the explosion. That change in the slope of the light
curve differentiates between a PM central engine and a collapsar. Our
axisymmetric models produce insufficient amounts of Ni during the
early post-bounce phase to make the radioative luminosity large enough
to be compatible with that of a SLSN. Indeed, the decay of the amount
of Ni existing at $t=t_{\rm f}$ yields peak bolometric luminosities
$\lesssim \zehn{41}\,\ergs$. Taking aside the contribution of the
central engine, the rough estimates of the possible amount of Ni
synthesized in our models, yields that magnetorotational explosions
lead to brighter SNe than neutrino-driven ones. However, there are
other conceivable sources of Ni that our models do not include (e.g.,
winds emerging from a collapsar accretion disk). Since the luminosity
powered by nuclear activity is $L_{\rm nuc}\propto M_{\rm Ni}$, we
cannot disregard the hypothesis that our stellar progenitors could
produce even larger radioactive peak luminosity values. We will
explore in a subsequent publication whether 3D versions of our models
yield larger Ni masses.

To compute the light curves, we assume that the central engine injects
a fraction of its rotational energy into the ejecta on the typical
spin-down timescale of a NS ($t_{\rm pm}$) or of a BH ($\sim t_{\rm
  BZ}$). The large difference between the engine injection timescale
and the effective photon diffusion time ($t_{\rm r}$) does not help to
maximize the peak luminosity of the optical transient
\citep{Arnett__1982__apj__TypeIsupernovaeI-Analyticsolutionsfortheearlypartofthelightcurve}. Nevertheless,
the luminosity reaches values within the realm of SLSNe or HNe.  The
estimation of the spin-down timescale is, however, only
approximate. We have computed from our models the values of the
instantaneous timescale for the loss of rotational energy of the PNS,
$\tau_{\rm rot}=\Erotpns/|d\Erotpns/dt|$.  In line with the results of
\citetalias{Aloy__2021__MonthlyNoticesoftheRoyalAstronomicalSociety__MagnetorotationalCoreCollapseofPossibleGRBProgenitorsII.FormationofProtomagnetarsandCollapsars},
we find that $\tau_{\rm rot}$ varies from seconds to hundreds of
seconds, thus, $\tau_{\rm rot}\ll t_{\rm pm}$. Nevertheless, in our NS
forming models, the Kelvin-Helmhotz phase has not finished by the end
of the computed time. Hence, we take $\tau_{\rm rot}$ as a lower bound
of $t_{\rm pm}$. For BH forming models, our code cannot provide a
significant insight on $t_{\rm BZ}$, since our simulations end by the
birth of the BH, but previous analytic work suggest $t_{\rm BZ}\propto
B_{\rm bh}^{-2}M_{\rm bh}^{-1}$. That dependence results in $t_{\rm
  BZ}\gg t_{\rm pm}$, which brings the spin-down time of the BH closer
to $t_{\rm r}$, boosting the contribution of the central engine to the
luminosity in BH forming cases compared to NS forming ones.We have
estimated the luminosity without accounting for the contribution of
the ejecta/circumstellar medium interaction, something beyond the
scope of this paper. But, we envisage that the large energy input
coming from the central engine may enhance the aforementioned
interaction, especially in BH forming cases. Hence, we deem as not
unlikely that our bolometric luminosity estimates are lower bounds of
more comprehensive and explicit calculations

Our results confirm that a subset of the massive, rapidly rotating and
magnetized stars of \AguiDet that undergo CHE yield SN explosions with
diagnostic explosion energies of the order of that in ordinary
CCSNe. Since the stellar progenitors have lost their H envelope and
most of the He, the explosion of these models results into Type Ic
SNe.  The former models are optimally suited to produce the central
engine of high energy transients such as GRBs because they meet three
conditions. First, the compact remnant left after core bounce
possesses a large rotational energy (typically above
$10^{52}\,\erg$). Second, the stellar progenitor possesses seed
magnetic fields that are amplified during collapse. These fields may
facilitate the transfer of the compact core rotational energy to the
surrounding matter. Third, a fraction of the matter surrounding the
central object has a specific angular momentum large enough to support
matter against direct collapse, hence enabling the possibility of
forming an accretion disc. Whether the central engine is a collapsar
or a PM depends on the lifetime of the PNS formed after core collapse.

\section{Acknowledgements}
\label{Sek:Ackno}

This work has been supported by the Spanish Ministry of Science,
Education and Universities (PGC2018-095984-B-I00) and the Valencian
Community (PROMETEU/2019/071). We furthermore thank for support from
the COST Actions PHAROS CA16214 and GWverse CA16104. MO acknowledges
support from the European Research Council under grant
EUROPIUM-667912, and from the the Deutsche Forschungsgemeinschaft
(DFG, German Research Foundation) - Projektnummer 279384907 - SFB 1245
as well as from the Spanish Ministry of Science via the Ram\'on y
Cajal programme (RYC2018-024938-I). The authors thankfully acknowledge
the computer resources and the technical support provided by grants
AECT-2018-3-0010, AECT-2019-1-0009, AECT-2020-3-0005, and
AECT-2021-1-0004 of the Spanish Supercomputing Network on cluster
MareNostrum of the Barcelona Supercomputing Centre - Centro Nacional
de Supercomputaci\'on, on clusters Tirant and Lluisvives of the Servei
d’Inform\`atica of the University of Valencia (financed by the FEDER
funds for Scientific Infrastructures; IDIFEDER-2018-063).

\section*{Data Availability}

The data underlying this article will be shared on reasonable request
to the corresponding authors.

\bsp	
\label{lastpage}

\end{document}